\documentclass[journal]{IEEEtran}

\usepackage{epsfig, amsmath, amssymb, cite, booktabs, array, multirow}
\usepackage{dblfloatfix} 
\usepackage{arydshln}
\usepackage{mathtools}
\usepackage{mathrsfs}
\usepackage{boldline}
\usepackage{algorithm}
\usepackage{algorithmic}
\usepackage{multirow}
\usepackage[normalem]{ulem}
\usepackage{graphicx}
\usepackage{enumitem}
\usepackage{color}
\useunder{\uline}{\ul}{}
\makeatletter
\newcommand*{\rom}[1]{\expandafter\@slowromancap\romannumeral #1@}
\makeatother

\graphicspath{{./figure/}}

\begin{document}

\title{TutorNet: Towards Flexible Knowledge Distillation for End-to-End Speech Recognition}

\author{Ji~Won~Yoon,~\IEEEmembership{Student Member,~IEEE}, Hyeonseung~Lee,~\IEEEmembership{Student Member,~IEEE},
Hyung~Yong~Kim,~\IEEEmembership{Student Member,~IEEE},
Won~Ik~Cho,~\IEEEmembership{Student Member,~IEEE}, and Nam~Soo~Kim,~\IEEEmembership{Senior Member,~IEEE}

\thanks{J. W. Yoon, H. S. Lee, H. Y. Kim, W. I. Cho, and N. S. Kim are with the Department of Electrical and Computer Engineering and the Institute of New Media and Communications, Seoul National University, Seoul, Korea (e-mail: jwyoon@hi.snu.ac.kr, hslee@hi.snu.ac.kr, hykim@hi.snu.ac.kr, wicho@hi.snu.ac.kr, nkim@snu.ac.kr) (Corresponding author: Nam Soo Kim).}
\thanks{© 2021 IEEE. Personal use of this material is permitted. Permission from
IEEE must be obtained for all other uses, in any current or future media,
including reprinting/republishing this material for advertising or promotional
purposes, creating new collective works, for resale or redistribution to servers
or lists, or reuse of any copyrighted component of this work in other works.}
\thanks{Digital Object Identifier 10.1109/TASLP.2021.3071662}}

\ifCLASSOPTIONpeerreview
\author{\IEEEauthorblockN{Ji~Won~Yoon,~\IEEEmembership{Student Member,~IEEE} and Nam~Soo~Kim,~\IEEEmembership{Senior Member,~IEEE}}
\IEEEauthorblockA{Department of Electrical and Computer Engineering and INMC,\\
Seoul National University \\
1 Gwanak-ro, Gwanak-gu, Seoul 08826, Korea \\
Tel: +82-2-880-8439 Fax: +82-2-880-8219 E-mail:
nkim@snu.ac.kr}} \fi


\maketitle

\begin{abstract}
    In recent years, there has been a great deal of research in developing end-to-end speech recognition models, which enable simplifying the traditional pipeline and achieving promising results. Despite their remarkable performance improvements, end-to-end models typically require expensive computational cost to show successful performance.
    To reduce this computational burden, knowledge distillation (KD), which is a popular model compression method, has been used to transfer knowledge from a deep and complex model (teacher) to a shallower and simpler model (student).
    Previous KD approaches have commonly designed the architecture of the student model by reducing the width per layer or the number of layers of the teacher model.
    This structural reduction scheme might limit the flexibility of model selection since the student model structure should be similar to that of the given teacher.
    To cope with this limitation, we propose a KD method for end-to-end speech recognition, namely TutorNet, that applies KD techniques across different types of neural networks at the hidden representation-level as well as the output-level.
    For concrete realizations, we firstly apply representation-level knowledge distillation (RKD) during the initialization step, and then apply the softmax-level knowledge distillation (SKD) combined with the original task learning.
    When the student is trained with RKD, we make use of frame weighting that points out the frames to which the teacher model pays more attention.
    Through a number of experiments on LibriSpeech dataset, it is verified that the proposed method not only distills the knowledge between networks with different topologies but also significantly contributes to improving the word error rate (WER) performance of the distilled student.
    Interestingly, TutorNet allows the student model to surpass its teacher's performance in some particular cases.
\end{abstract}


\IEEEpeerreviewmaketitle

\section{Introduction}

\IEEEPARstart{R}{ecently}, there has been a huge interest in the research on end-to-end speech recognition, such as the connectionist temporal classification (CTC) \cite{graves-et-al:scheme}, attention encoder-decoder (AED) \cite{chorowski-et-al:scheme}, and recurrent neural network transducer (RNN-T) \cite{graves-rnn-t:scheme}. End-to-end models directly map an input speech signal into the corresponding sequence of words yielding better performance compared to the conventional deep neural network (DNN)-hidden Markov model (HMM) hybrid systems. However, most end-to-end models require heavy computation and a large number of parameters for a successful performance. In order to achieve competency within the constraints on resources, it is desirable to design a more lightweight model.

\begin{figure}[t]
	{
		\begin{center}
			\begin{tabular}{c}
				\includegraphics[height=8cm]{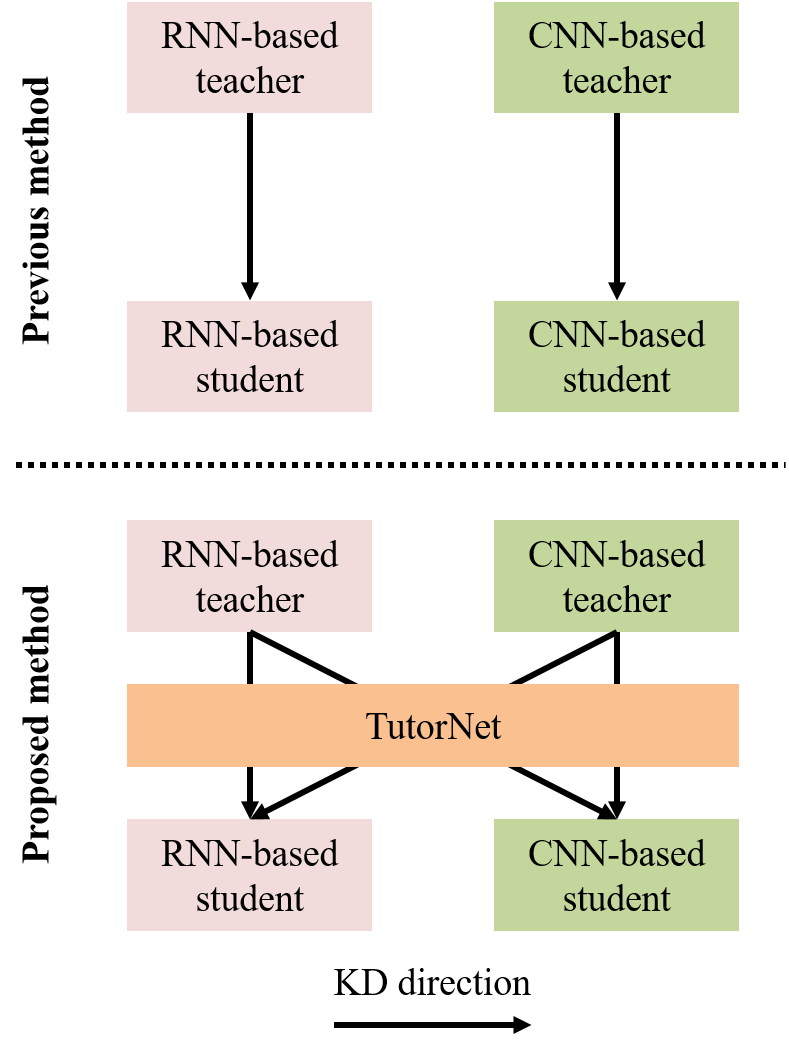}
			\end{tabular}
		\end{center}
	}
	\caption{Conceptual diagram of TutorNet. Prior approaches typically transfer knowledge between models with the same structure (i.e., just reducing the layer size or the width per layer). The proposed method can connect teacher/student models in the knowledge distillation task, in which both models have different topologies.}
	\label{diagram}
\end{figure}

Knowledge distillation (KD) is one of the most popular approaches for model compression, which aims at transferring knowledge from a bigger network (teacher) to a much smaller network (student). It is generally assumed that the teacher has been trained separately while consuming huge computation. The goal of KD is to make the student mimic the behavior of the teacher, leading to better performance compared to the case when it is solely trained.
In designing the student model's architecture, conventional KD approaches typically reduce the layer width or the number of layers of the teacher model, which reduces the number of parameters dramatically. However, this simple structure reduction scheme usually limits the flexibility of model selection. For speech recognition, there have been various types of end-to-end models, such as DeepSpeech2 \cite{amodei-et-al:scheme}, Wav2Letter \cite{collobert-et-al:scheme}, Jasper \cite{li-et-al:scheme}, QuartzNet \cite{quartznet:scheme}, AED \cite{chorowski-et-al:scheme}, and Transformer \cite{transformer:scheme}. 
Despite the diversity of such models available, the conventional KD technique usually employs the student model structure similar to that of the given teacher model. In other words, no matter how suitable some models are as a teacher (high performance) or a student (less parameter, fast inference, etc.), the adoption has often been restricted due to their structure being different from the counterpart. For example, Transformer has been proven to perform better in the speech area, but it has not been used as a teacher for the other models, such as CTC or AED models.

To handle this limitation, in this paper, we attempt to apply KD techniques to an unexplored setting where the architecture of the student is inherently different from that of the teacher, as conceptually displayed in Fig. \ref{diagram}.
For instance, via the proposed approach, a recurrent neural network (RNN)-based DeepSpeech2 can benefit from the distilled knowledge of a convolutional neural network (CNN)-based Jasper not only at the hidden representation-level but also at the output-level. The opposite case, i.e., transferring knowledge from CNN-based Jasper to RNN-based DeepSpeech2, is also possible. Furthermore, the student can be trained with the knowledge of both CNN-based Jasper and RNN-based DeepSpeech2. In addition to the CTC model, the proposed method can be applied to the other types of models.
We apply a network called TutorNet, which connects the teacher and student models even when the two models have different types of structures.
TutorNet consists of two stages: (1) representation-level KD (RKD) for initializing the network parameters and (2) softmax-level KD (SKD) for transferring softmax prediction, where both stages can be applied regardless of the difference in model architecture. When training the student model with RKD, we utilize frame weighting that picks the frames to which the teacher model pays attention.
We verify the effectiveness of the proposed method via a substantial model comparison.

To summarize, the main contributions of this paper are:
\begin{itemize}
    \item We introduce TutorNet for transferring the hidden representation and softmax values across different types of neural networks. On top of that, we also make use of frame weighting, reflecting which frames are important for KD.
    \item To distill frame-level posterior in the CTC framework, we suggest that $l_2$ loss is more suitable than the conventional Kullback-Leibler (KL)-divergence.
    \item The proposed method substantially outperformed the other conventional KD methods in several speech recognition experiments. It is noted that the student model performs even better than its teacher in some particular cases.
    \item TutorNet is applicable not only to the CTC-based model but also to the other end-to-end speech recognition models.
\end{itemize}

The rest of the paper is organized as follows: Related work is described in Section \rom{2}. We introduce the proposed KD method, namely TutorNet, in Section \rom{3}, describe experimental settings in Section \rom{4}, and then present experimental results obtained under various settings in Section \rom{5}. Finally, Section \rom{6} concludes the paper.

\section{Related work}
\subsection{Connectionist Temporal Classification (CTC)}

Generally, an end-to-end speech recognition model directly maps a sequence of input acoustic features $x_{1:T}=\{x_{1},...,x_{T}\}$ into a sequence of target labels $y_{1:N}=\{y_{1},...,y_{N}\}$ where $y_{n} \in \mathcal{Y}$ with $\mathcal{Y}$ being the set of labels in texts. $T$ and $N$ are respectively the total number of frames and the length of the target label sequence. 
To deal with the sequence-to-sequence mapping problem when the two sequences have unequal lengths, the CTC framework \cite{graves-et-al:scheme} introduces~``blank" as an additional label and allows the repetition of all labels across frames. An alignment sequence $\pi_{1:T}=\{\pi_{1},...,\pi_{T}\}$ is a sequence of initial output labels for each frame, as every input frame $x_{t}$ is mapped to a certain label $\pi_{t} \in \mathcal{Y'}$ where $\mathcal{Y'} = \mathcal{Y} \cup \{blank\}$. A mapping function $\mathcal{B}$, which is defined as $y = \mathcal{B}(\pi)$, converts the alignment sequence $\pi$ into the final output sequence $y$ after merging consecutive repeated characters and removing blank labels. 
For example, two alignment sequences $\{\varepsilon, c, c, c, \varepsilon, a, \varepsilon, \varepsilon, t, t, \varepsilon\}$ and $\{c, c, \varepsilon, \varepsilon, a, a, \varepsilon, \varepsilon, \varepsilon, \varepsilon, t\}$ (using ~`$\varepsilon$' to denote blank label) correspond to the same sequence $\{c, a, t\}$ through the mapping function $\mathcal{B}$. The alignment between the input $x$ and the target output $y$ is not explicitly required in CTC training.
The conditional probability of the target label sequence $y$ given the input sequence $x$ is defined as
\begin{equation}
   p(y \vert x) = \sum_{\pi \in \mathcal{B}^{-1}(y)} p(\pi \vert x).
\end{equation}
where $\mathcal{B}^{-1}$ denotes the inverse mapping and returns all possible alignment sequences compatible with $y$.
 The conditional probability of a path ${\pi}=\{\pi_{1},...,\pi_{T}\}$ can be calculated as follows:
 \begin{equation}
   p(\pi \vert x) = \prod_{t=1}^{T} p(\pi_{t} \vert x).
\end{equation}
 Given the target $y$ and the input $x$, the loss function $\mathcal{L}_{CTC}$ is defined as
\begin{equation}
\label{ctc_loss}
   \mathcal{L}_{CTC} = - \sum_{(x,y) \in Z} \ln p(y \vert x) 
\end{equation}%
where $Z$ represents a training dataset. 
\begin{figure*}[t]
	{
		\begin{center}
			\begin{tabular}{c}
				\includegraphics[height=7.2cm]{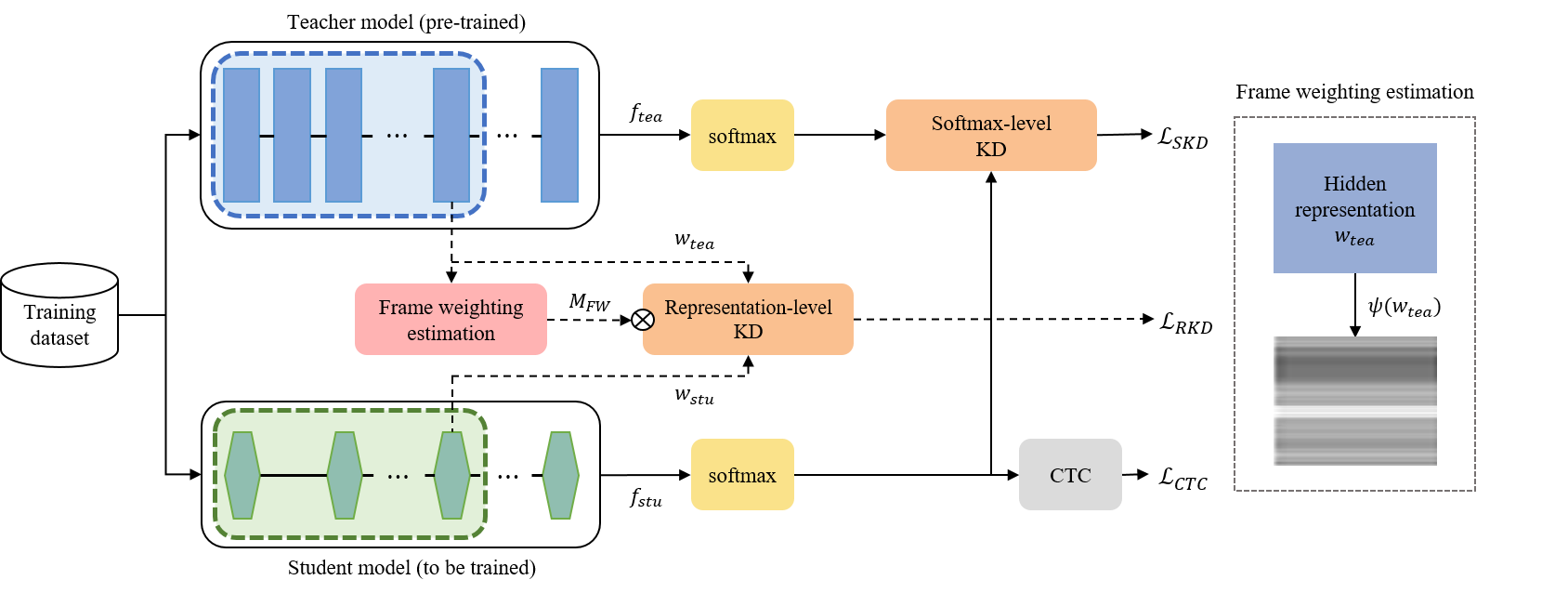}
			\end{tabular}
		\end{center}
	}
	\caption{An overview of our KD approach for end-to-end speech recognition. There are two stages of training: (1) representation-level KD with frame weighting and (2) softmax-level KD. Firstly, the process of representation-level KD initializes the student by minimizing the distance between hidden representation of the teacher model $w_{tea}$ and student model $w_{stu}$. In order to consider frame weighting, reflecting which frames are important for KD, it takes the hidden representation of the teacher model $w_{tea}$ as an input and yields frame weighting mask $M_{FW}$. Afterwards, the student model, initialized via representation-level KD, is trained with the normal CTC training and softmax-level KD.}
	\label{overview}
\end{figure*}

\subsection{Attention-based Encoder-Decoder (AED)}
An alternative approach to the end-to-end mapping between speech and
label sequences is to use the attention-based model. Unlike the CTC framework, AED directly predicts each target without requiring any alignment sequence. The framework consists of two sub-modules $Encoder$ and $AttentionDecoder$, so that it can learn two different lengths of sequences based on the cross-entropy criterion. The model predicts the posterior probability of the output transcription given the input speech features as follows:
\begin{equation}
    \label{attention_encoder}
    h = Encoder(x)
\end{equation}
\begin{equation}
    \label{attention_decoder}
    p(y_{u} \vert x, y_{1:u-1}) = AttentionDecoder(h, y_{1:u-1}).
\end{equation}
where $x$ and $h$ are sequences of input speech features and encoded vectors respectively, and $y$ is a sequence of output text units whose length is $U$. 
$Encoder$ extracts encoded vectors $h$ from the input speech features $x$ in (\ref{attention_encoder}). Then $AttentionDecoder$ network predicts
the next output symbol conditioned on the full sequence of previous
predictions and acoustics, which can be defined as $p(y_{u} \vert x, y_{1:u-1})$. The attention mechanism determines which encoded vectors should be attended in order to predict the next output symbol.

\subsection{Knowledge Distillation}
Neural network models usually perform well with a large number of parameters. However, as a model architecture gets deeper, it requires heavy computation for both training and testing. To mitigate this computational burden, there has been a long line of research on KD, which aims at distilling knowledge from a big teacher model to a small student model. With this additional transfer procedure, the student can perform better compared to naive training. Existing KD methods typically fall into two categories: (1) transferring class probability and (2) transferring the representation of the hidden layer.

Generally, the output layer for the classification tasks uses softmax as an activation function. The output of the teacher model is a probability distribution over the target classes, and the sum of the outputs equals 1. Hinton \textit{et al.} \cite{hinton_kd-et-al:scheme} first introduced KD, which distills class probability by minimizing the KL-divergence between the softmax outputs of the teacher and student. Compared to the one-hot label, the teacher's softmax prediction has a nonzero probability value for each target class. This soft label is normally considered more informative than the one-hot encoded ground truth, further improving the student model in KD.
The KD technique mentioned above only considers the output of the teacher model. In the case of transferring the hidden representation, some KD methods \cite{dodeep:scheme,romero-et-al:scheme,at:scheme,fsp:scheme,jacobian:scheme}, especially in image processing, proposed transferring the representation-level information of the hidden layers where the mean squared error (MSE) between the representation-level knowledge of both models is minimized.

For speech, Li \textit{et al.} \cite{firstasr:scheme} first attempted to apply the teacher-student learning. In previous studies for speech recognition, KD typically has been applied to DNN-HMM hybrid systems by minimizing the frame-level cross-entropy loss between the output distributions of the teacher and student \cite{chebotar-et-al:scheme,watanabe-et-al:scheme,lu-et-al:scheme,fukuda-et-al:scheme}.
Geras \textit{et al.} \cite{blending:scheme} proposed KD method that distills the softmax-level knowledge from RNN-based model to CNN-based model in the DNN-HMM framework. Wong \textit{et al.} \cite{seq:scheme, seq2:scheme} investigated sequence-level knowledge distillation to DNN-HMM trained by sequence discriminative criteria. In the case of KD under the cross-entropy criteria, the KD loss can be calculated as
\begin{equation} 
\mathcal{L}_{KD}=-\sum_{y} p_{tea}(y \vert x) \ln p_{stu}(y \vert x)
\end{equation}
where $p_{tea}(y \vert x) $ represents the posterior probability of the target label $y$
given the input $x$ yielded by the teacher model, and $p_{stu}(y \vert x)$  is that of the
student model.

The same frame-level KD method has also been applied to KD of CTC models \cite{senior-et-al:scheme}. The frame-level KD in CTC framework can be computed as follows:
\begin{equation} 
\label{frame_kd_loss}
\mathcal{L}_{CTC-{KD_{frame}}}=-\sum_{x\in Z} \sum^{T}_{t=1} \sum_{k \in \mathcal{Y'}} p_{tea}(k \vert x_t) \ln p_{stu}(k \vert x_t)
\end{equation}
where $p_{tea}(k \vert x_t)$ and $p_{stu}(k \vert x_t)$ denote the posterior probability of the teacher and student CTC models, respectively.
However, as reported in previous studies \cite{senior-et-al:scheme,takashima-et-al:scheme, takashima-et-al2:scheme}, applying the frame-level KD approach to the CTC-based speech recognition system can worsen the word error rate (WER) performance compared with the CTC model, which is trained only with the ground truth.

To address this problem, Takashima \textit{et al.} \cite{takashima-et-al:scheme} proposed a KD method that distills a sequence-level knowledge in the CTC framework using the teacher model's N-best hypotheses. Kurata and Audhkhasi \cite{kurata2-et-al:scheme,kurata-et-al:scheme} also introduced a KD approach for long short-term memory (LSTM)-CTC-based speech recognition model where a student can be trained using the frame-wise alignment of the teacher.

\section{Proposed Method}

Our main goal is to transfer the teacher model's knowledge to the student model without being restricted to the types of model architecture. As shown in Fig. \ref{overview}, TutorNet is mainly composed of two stages: (1) representation-level KD (RKD) with frame weighting and (2) softmax-level KD (SKD).
In Section \rom{3}-A, we first describe the initialization step RKD, where the student benefits from the hidden representations of the teacher.
Although the two models have different types of architecture, the proposed method enables flexible knowledge transfer at the representation-level.
While training the student with RKD, we make use of frame weighting that picks the frames to which the teacher model pays attention.
Section \rom{3}-B subsequently introduces SKD, which allows the student to frame-wisely track the posterior distribution of the teacher.
Previous studies \cite{senior-et-al:scheme, takashima-et-al:scheme, takashima-et-al2:scheme} have found that applying the conventional frame-level KD to the student CTC model is a challenging problem.
Based on these observations, we adopt the $l_2$ loss function instead of the conventional objective function.

\subsection{Representation-Level Knowledge Distillation with Frame Weighting}
As mentioned above, the most frequently employed KD approach for speech recognition is to train a student with the teacher's softmax prediction as a target, besides the one-hot encoded ground truth. However, the hidden representations of the teacher model are also considered important to provide essential knowledge for training the other.
Moreover, if we can transfer hidden representations regardless of the types of neural network architecture, more flexible and effective distillation will be possible. 

\subsubsection{Hidden representation matching using 1D convolutional layer}
Let $w_{tea}^{(i)}$ and $w_{stu}^{(j)}$ respectively denote the hidden representation from the \textit{i}-th and \textit{j}-th layers of the teacher and student models where both models are assumed to have different architecture.
In the CTC framework, when the speech signal $x$ is given as an input to both models, the \textit{i}-th hidden layer representation of the teacher model can be expressed as $w_{tea}^{(i)}(x) \in R^{T \times D_{t}}$, where $T$ represents the total number of frames with the hidden layer dimension $D_{t}$. In a similar way, $w_{stu}^{(j)}(x) \in R^{T \times D_{s}}$, where $D_{s}$ denotes the hidden layer width.
Since usually the hidden layer dimensions $D_{t}$ and $D_{s}$ are different, we apply a convolutional layer to minimize the following mismatch error:
\begin{equation}
\label{hidden}
   \Vert{w_{tea}^{(i)}(x)-c_{\theta}(w_{stu}^{(j)}(x))}\Vert_2^2
\end{equation}%
where $c_{\theta}$ is a 1D convolutional layer, and $\theta$ denotes its parameters. The essential difference from Fitnets \cite{romero-et-al:scheme} is that we attempt to transfer the hidden representation across different types of neural networks for end-to-end speech recognition, which is an unexplored area in the related research.
Concerning that $w_{stu}$ has a different structural nature from $w_{tea}$, the convolutional layer $c_{\theta}$ not only converts the hidden layer size of the student from $D_{s}$ to $D_{t}$ but also effectively distills the teacher's hidden layer information even when both models have different architecture.

\begin{figure}[t]
	{
		\begin{center}
			\begin{tabular}{c}
				\includegraphics[height=2.75cm]{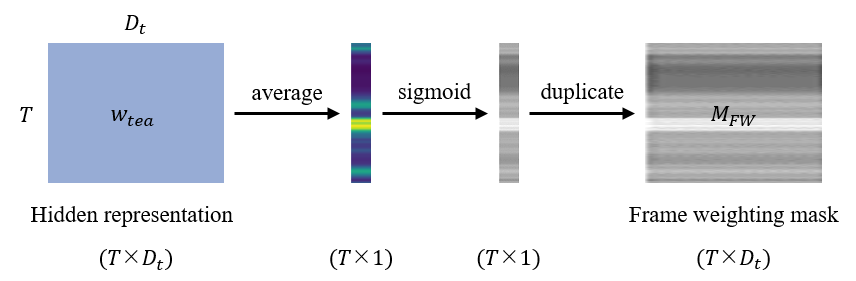}
			\end{tabular}
		\end{center}
	}
	\caption{The procedure for computing the frame weighting mask $M_{FW}$.}
	\label{frame_weighting}
\end{figure}

\subsubsection{Frame weighting}
It is generally accepted that each frame of the speech has different importance for KD. For instance, active speech periods should be treated more importantly than the silence periods. For this reason, instead of transferring all the hidden representations equally, we employ frame weighting that puts more emphasis on frames that correspond to the neurons with high activations.
A frame weighting function $\psi(\cdot)$ takes the teacher's representation $w_{tea}\in R^{T \times D_{t}}$ as an input and outputs a frame weighting mask $M_{FW}\in R^{T \times D_{t}}$. As illustrated in Fig. \ref{frame_weighting}, the procedure for computing $M_{FW}$ is as follows:
\begin{enumerate}[label=\alph*)]
    \item Average $w_{tea}$ over the hidden representation dimension (horizontal) axis, where the resulting vector is in $R^{T \times 1}$.
    \item For normalization, map the average values to the interval $[0, 1]$ by using a sigmoid function $\sigma(\cdot)$.
    \item Replicate the values of the vector for $D_{t}$ times along the hidden layer dimension to produce $M_{FW}$.

\end{enumerate}

RKD is used to initialize the student's parameters before CTC training. It aims to transfer the hidden representations which serve good initialization of the model parameters.
In RKD, considering the frame weighting mask $M_{FW}$, the student model is trained to minimize the following loss function:
\begin{equation}
   \label{rkd}
   \mathcal{L}_{RKD}=\sum_{x\in Z}\sum_{(i,j)\in\mathcal{I}}\Vert{M_{FW}\odot(w_{tea}^{(i)}(x)-c_{\theta}(w_{stu}^{(j)}(x))})\Vert_2^2
\end{equation}%
where $\mathcal{I}$ represents the set of candidate layer index pairs, and $\odot$ indicates the Hadamard product. In (\ref{rkd}), it is assumed that the \textit{i}-th layer of the teacher model is transferred to the \textit{j}-th layer of the student when $(i,j)\in\mathcal{I}$.

\subsection{Softmax-Level Knowledge Distillation}
\begin{figure}[t]
	{
		\begin{center}
			\begin{tabular}{c}
				\includegraphics[height=3.5cm]{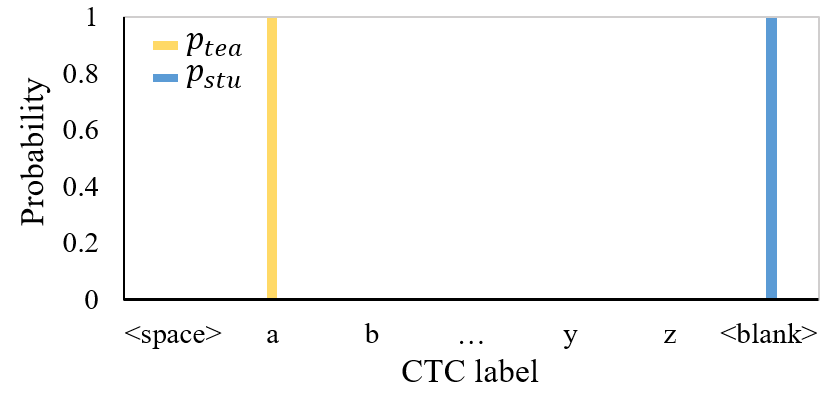}
			\end{tabular}
		\end{center}
	}
	\caption{Given two distributions $p_{tea}$ and $p_{stu}$ of a certain frame $t$, there is no overlap between the two distributions.}
	\label{ctc}
\end{figure}
Before we describe SKD for distilling frame-level posterior in the CTC framework, it might be beneficial to review some characteristics of the CTC model.
The output of the CTC model has two notable characteristics to consider when transferring a posterior distribution to the other. 
First, as noted in prior studies \cite{graves-et-al:scheme, sak2015learning}, the softmax prediction obtained from a CTC-trained model is very spiky. This indicates that the softmax output tends to be similar to the one-hot vector.
Secondly, since CTC is an alignment-free framework, CTC models trained with the same training data can have different frame-level alignments.
In other words, the frame-level alignment yielded by the student can be different from that of the teacher.
Unfortunately, this characteristic makes KD difficult because the KL-divergence and cross-entropy can hardly converge.
Suppose we have two probability distributions $p_{tea}$ and $p_{stu}$ obtained at a certain frame $t$. As shown in Fig. \ref{ctc}, $p_{tea}$ has the highest probability on label~`a' and $p_{stu}$ on~`blank'.
In this case, the KL-divergence becomes infinity and KD is hard to converge.

In previous studies \cite{senior-et-al:scheme,takashima-et-al:scheme, takashima-et-al2:scheme}, it has been confirmed that applying the conventional frame-level KD approach can worsen the performance of a student CTC model compared with a model trained without KD.
Also, we tried to train a student CTC model with the interpolation between the original CTC loss in (\ref{ctc_loss}) and frame-level KD in (\ref{frame_kd_loss}), but it failed to converge as reported in \cite{takashima-et-al:scheme}.

To deal with this instability problem, we propose to use $l_2$ loss instead of the KL-divergence. Since $l_2$ distance between two distributions is always bounded unlike the KL-divergence, it improves the numerical stability of the distillation loss.
This alternative approach, namely SKD, allows student to stably learn the alignment of all the output labels, including blank ones. The SKD loss $\mathcal{L}_{SKD}$ to train the student model is given as
\begin{equation}
\mathcal{L}_{SKD}=\sum_{x\in Z}
\Vert{softmax({f_{tea}(x)\over\tau})-softmax({f_{stu}(x)\over\tau})} \Vert_2^2
\end{equation}
where $f_{tea}$ and $f_{stu}$ are the logits obtained from the teacher and student models respectively, and $\tau$ represents a temperature parameter.
During SKD training, the SKD loss $L_{SKD}$ and the standard CTC loss $L_{CTC}$ are combined as an integrated loss.
Thus, the final objective function for SKD can be formulated as

\begin{equation}
\label{skd_ctc}
\mathcal{L}_{CTC-SKD}=L_{CTC}+ \lambda_{SKD} \cdot L_{SKD}
\end{equation}%
where $\lambda_{SKD}$ is a tunable parameter.

\subsection{Learning Procedure}
Our approach includes two stages of training: RKD as the initialization step and SKD in conjunction with the CTC objective as the fine-tuning step.
Firstly, by minimizing $L_{RKD}$ without the CTC loss function, the student can mimic the teacher's behavior at the representation-level. To effectively transfer the representation-level knowledge of the teacher, we apply the frame weighting mask $M_{FW}$ as in (\ref{rkd}). 
After the RKD initialization, the student model is trained with SKD. As the student is trained to minimize the combined loss $L_{CTC-SKD}$, it can frame-wisely track the softmax prediction of the teacher.

\section{Experimental settings}
\subsection{Dataset} We evaluated the performance of the proposed method on two speech datasets: LibriSpeech \cite{panayotov-et-al:scheme} and AISHELL-2 \cite{aishell:scheme}. LibriSpeech is a large-scale (about 1000 hours) English speech corpus derived from audiobooks, sampled at 16kHz. The dataset is divided into \textit{clean} and \textit{other}. In the experiment,~``train-clean-100",~``train-clean-360", and~``train-other-500" were used in the training phase. For evaluation, ~``dev-clean",~``dev-other",~``test-clean", and ~``test-other" were applied. We also conducted our experiments on AISHELL-2, the Mandarin read-speech dataset (around 1000 hours of training data). The model evaluated over AISHELL2-2018A-EVAL data, containing dev set (2500 utterances from 5 speakers) and test set (5000 utterances from 10 speakers).

\subsection{Performance Metrics} For LibriSpeech, we measured two metrics: word error rate (WER) and relative error rate reduction (RERR). WER is commonly employed to quantify speech recognition performance. To calculate WER, the number of errors is obtained by counting the substitutions, insertions, and deletions that occur in the recognition result. Then, it is divided by the total number of words in the correct sentence.
RERR shows how much the WER is reduced, in proportion, compared to the baseline. For the Mandarin dataset, we measured the character error rate (CER) instead of WER. This is because a single character often represents a word for the Mandarin writing system. CER follows the same formula of WER, but with characters as the unit.

\subsection{Model Configuration} For KD, we adopted the following different speech recognition models which were used as the teacher or student model: 
\begin{itemize}
    \item CTC model
        \begin{itemize}
        \item Jasper Dense Residual (Jasper DR) \cite{li-et-al:scheme}: Jasper DR is a deep time-delay neural network (TDNN) composed of blocks of 1D convolutional layers. This differs from the original  Jasper in that it has dense residual connections. The output of the convolution block is fed as an input to all the blocks via dense residual connections. In our experiments, we applied pre-trained Jasper DR, which consists of 54 convolutional layers, as the CNN-based teacher model.
        \item DeepSpeech2 \cite{amodei-et-al:scheme}: DeepSpeech2 has the architecture of a deep RNN with a combination of convolutional and fully connected layers. As for the RNN-based model, we applied DeepSpeech2, consisting of two 2D convolutional layers followed by three 512-dimensional bidirectional LSTM (BLSTM) layers and one fully connected layer.
        \item Jasper Mini: Jasper Mini is composed of blocks of depthwise separable 1D convolutional layers. Depthwise separable convolution reduces the number of parameters and the computation required in the convolutional operations. The main structural difference from the Jasper DR is that Jasper Mini consists of depthwise separable convolutions with no dense residual connection.
        As for the CNN-based student model, we adopted Jasper Mini with 33 depthwise separable convolutional layers.
        \item QuartzNet \cite{quartznet:scheme}: QuartzNet is composed of multiple blocks with residual connections between them. Each block consists of one or more modules with 1D time-channel separable convolutional layers, batch normalization, and ReLU. We adopted Quartznet, which consists of 54 1D time-channel separable convolutional layers.
        \end{itemize}

    \item Hybrid CTC/attention model \cite{hybrid_ctc_attention:scheme}
    \begin{itemize}
        \item Attention-based encoder-decoder (AED): Motivated by the prior studies \cite{vgg-asr:scheme, hori-et-al:scheme}, the encoder contained the initial layers of the VGG net architecture \cite{vgg:scheme} followed by four 1024-dimensional BLSTM layers and a linear projection layer, which yields better performance than the pyramid BLSTM \cite{las:scheme} in many cases. 
        We used a location-based attention mechanism, and the decoder network was a 2-layer LSTM with 1024 cells.
        The model was trained by both CTC and attention model objectives simultaneously.
    \end{itemize}
    
\end{itemize}
Table \ref{baseline} compares WERs obtained from the three baseline models when greedy decoding was applied and the number of parameters. Since the DeepSpeech2 baseline with BPE units failed to converge with a single BPE CTC loss, we adopted the CTC model, initialized with RKD over the first 100 steps, as the student baseline.

\begin{table}[t]
\centering
\caption{Comparison across baseline model configurations when greedy decoding was applied. For AISHELL-2, we measured CER (\%) instead of WER (\%).}
\label{baseline}
\begin{tabular}{cccccc}
\toprule
Dataset & Model Type & Unit Set &  WER (\%) & Params. (M)\\
\midrule
& Jasper DR & Character & 3.61 & 332.63\\
& DeepSpeech2 & Character  & 7.64 & 13.19\\
& Jasper Mini & Character  & 8.66 & 8.19\\
LibriSpeech & QuartzNet & Character & 5.66 & 13.09\\
& AED & BPE & 5.80 & 174.10\\
& Transformer & BPE & 3.21 & 78.36\\
& DeepSpeech2 & BPE & 8.82 & 21.71\\
\midrule
& Jasper DR & Character & 9.69 & 337.94\\
AISHELL-2& Jasper Mini & Character & 11.77 & 13.50\\
& DeepSpeech2 & Character & 13.40 & 15.77\\
\bottomrule
\end{tabular}
\end{table}
\subsection{Implementation Details} Our experiments were conducted using three toolkits: OpenSeq2Seq \cite{openseq2seq}, ESPnet \cite{espnet:scheme}, and NeMo \cite{nemo:scheme}. When training the CTC-based model, we mainly used OpenSeq2Seq. In the case of hybrid CTC/attention model, such as AED and Transformer, we applied the ESPnet toolkit. Pre-trained Jasper DR for the Mandarin dataset was provided by the NeMo toolkit.
\begin{itemize}{
\item LibriSpeech
\begin{itemize}{
\item CTC model: We extracted 64-dimensional log-Mel filterbank features as the input, and the character set has a total of 29 labels with all lowercased Latin alphabet letters (a-z), apostrophe, space, and blank label. In the case of Jasper DR, we used the checkpoint provided by the OpenSeq2Seq toolkit. For character-level CTC-based student models, all the training excluding QuartzNet was performed on three Titan V GPUs, each with 12GB of memory. While training the RNN-based DeepSpeech2, Adam algorithm \cite{kingma-et-al:scheme} was employed as an optimizer with an initial learning rate of 0.001 that was reduced with polynomial decay. In the case of the CNN-based Jasper Mini, we used NovoGrad optimizer \cite{ginsburg-et-al:scheme} based on stochastic gradient descent (SGD). The initial learning rate of the optimizer started from 0.02, and it also had the same decaying policy as above. 
For RKD, we selected the last layer\footnote{CTC is regarded as a frame-level classification. Based on the general use of the last layer in KD of classification task, we chose the last layer for RKD training.} of the teacher and student models to transfer the hidden representation, and the student was trained for 5 epochs. When transferring the knowledge from CNN-based Jasper DR to RNN-based DeepSpeech2, the kernel size of $c_{\theta}$, $D_{t}$, and $D_{s}$ were 1, 1024, and 512, respectively. In distilling from RNN-based DeepSpeech2 to CNN-based Jasper Mini, the kernel size of $c_{\theta}$, $D_{t}$, and $D_{s}$ were 1, 512, and 1024, respectively.
After the RKD initialization, 50 epochs were spent for CTC training with SKD. In the case of QuartzNet, we trained the student model with 10 epochs for RKD initialization and 100 epochs for SKD training. The training of Quartznet was performed on three Titan RTX GPUs, each with 24GB of memory. In the case of QuartzNet, $D_{t}$, and $D_{s}$ were equal to 1024. We experimentally set the tunable parameter $\lambda_{SKD}$ to 0.25, which showed the best performance in the dev-clean set.  When applying beam-search decoding with language model (LM), we used KenLM \cite{Heafield2011KenLMFA} for 4-gram LM, where the LM weight, the word insertion weight, and the beam width were experimentally set to 2.0, 1.5, and 256, respectively. When training the CTC model distilled from Transformer, since we used the byte-pair encoding (BPE) \cite{bpe:scheme} method to construct the subword units (around 5000 tokens) for Transformer, the CTC-based student model was also trained with the same BPE units of the teacher model. The training was performed on three Titan RTX GPUs, each with 24GB of memory. In distilling the knowledge from Transformer to CTC-based model, 5 epochs and 50 epochs were spent for RKD and SKD, respectively.
\item Hybrid CTC/attention model: When training AED, we used one Titan V GPU. Adadelta \cite{adadelta:scheme} was employed as an optimizer with an initial learning rate of 1.0. During training and inference, we set the CTC weight to 0.5. For RKD, the student was trained for 1 epoch, and then it was trained with SKD for 10 epochs. In the case of Transformer, we used the pre-trained model provided by the ESPnet toolkit. We used BPE to construct the subword units for the label of both AED and Transformer (around 5000 tokens). For the external LM, we used pre-trained RNNLM from ESPnet toolkit. The beam size was experimentally set to 20. When transferring the knowledge from Transformer to CTC-based model, $D_{t}$, and $D_{s}$ were equal to 512. In KD from Transformer to AED model, $D_{t}$, and $D_{s}$ were 512 and 1024, respectively.
}\end{itemize}
\item AISHELL-2: For the Mandarin dataset, the character set has a total of 5207 labels. In the case of the teacher model, we used pre-trained Jasper DR, which was provided by the NeMo toolkit. When training the student model, we used three Titan V GPUs. For the CNN-based student model, the NovoGrad optimizer was adopted. In the case of the RNN-based student model, the Adam algorithm was employed as an optimizer. For RKD and SKD, 2 epochs and 20 epochs were spent, respectively. When training with SKD, we set $\lambda_{SKD}$ to 1.
}\end{itemize}

\subsection{Conventional KD Techniques for Performance Comparison} We applied the following conventional KD techniques for performance comparison\footnote{We tried to train a student CTC model with the interpolation between the original CTC loss in (\ref{ctc_loss}) and frame-level KD in (\ref{frame_kd_loss}), and it failed to converge as reported in \cite{takashima-et-al:scheme}. Therefore, the frame-level KD was not considered for comparison.}:
\begin{itemize}
    \item CTC baseline \cite{graves-jaitly:kl-one}: The student model is trained solely based on the ground truth as the target, i.e., $\lambda_{SKD}=0$ in (\ref{skd_ctc}).
    \item Sequence-level knowledge distillation \cite{takashima-et-al:scheme}: N-best hypotheses of the teacher are used as a sequence-level knowledge for KD.
    In this experiment, the 5-best\footnote{The Sequence-level KD criteria can be summarized as the weighted mean of the original CTC loss regarding each hypothesis of the label sequence. The posterior probabilities of the hypotheses estimated by the teacher CTC model are used as the weights of each CTC loss. In our experiments, most probabilities were concentrated in the highest-scoring hypothesis, which was at the top of the beam.  Even when we extracted more sentences than 5-best, the posterior probabilities were mostly concentrated on very few sentences. That is why we used 5-best hypotheses for the comparison.} hypotheses were extracted using KenLM, where the LM weight, the word insertion weight, and the beam width were experimentally set to 2.0, 1.5, and 256, respectively. 
    \item Guided CTC training \cite{kurata-et-al:scheme}: From the posterior distribution of the teacher, guided CTC training makes a mask that sets 1 only at the output label of the highest posterior at each frame. The student can be guided to align with the frame-level alignment of the teacher by using the guided mask.
\end{itemize}

\section{Experimental results}
\subsection{LibriSpeech}
For the convenience of notation and ease of comparison, we let CNN$_{DR}$, RNN$_{DS}$, and CNN$_{Mini}$ denote Jasper DR, DeepSpeech2, and Jasper Mini, respectively.
In the subsequent part of this paper, $A$ $\rightarrow$ $B$ means that model $A$ transfers knowledge to model $B$ where the former is a teacher and the latter is a student.
To verify the effectiveness of the proposed method in various situations, we tried six different transfer scenarios on LibriSpeech dataset:
\begin{itemize}
    \item CNN$_{DR}$ $\rightarrow$ RNN$_{DS}$: From CNN-based model (Jasper DR) to RNN-based model (DeepSpeech2)
    \item CNN$_{DR}$ $\rightarrow$ CNN$_{Mini}$: From CNN-based model (Jasper DR) to CNN-based model (Jasper Mini)
    \item RNN$_{DS}$ $\rightarrow$ CNN$_{Mini}$: From RNN-based model (DeepSpeech2) to CNN-based model (Jasper Mini)
    \item RNN$_{DS}$ $\rightarrow$ RNN$_{DS}$: From RNN-based model (DeepSpeech2) to RNN-based model (DeepSpeech2)
    \item (CNN$_{DR}$ \& RNN$_{DS}$) $\rightarrow$ CNN$_{Mini}$: From two teachers (Jasper DR \& DeepSpeech2) to CNN-based model (Jasper Mini)
    \item (CNN$_{DR}$ \& RNN$_{DS}$) $\rightarrow$ QuartzNet: From two teachers (Jasper DR \& DeepSpeech2) to CNN-based model (QuartzNet)
\end{itemize}
\subsubsection{CNN$_{DR}$ $\rightarrow$ RNN$_{DS}$}

\begin{table*}[t]
\centering
\caption{Comparison of WER (\%) on LibriSpeech.~``Ours" denotes TutorNet with SKD and RKD. The best result is in bold.}
\label{totalwer}
\begin{tabular}{@{}clcccccccc@{}}
\toprule
& & \multicolumn{4}{c}{WER (\%) w/o LM} & \multicolumn{4}{c}{WER (\%) w/ LM} \\ \cmidrule(l){3-10} 
KD & \hspace{15mm}Model & \multicolumn{2}{c}{dev} & \multicolumn{2}{c}{test} & \multicolumn{2}{c}{dev} & \multicolumn{2}{c}{test} \\ \cmidrule(l){3-10} 
& & clean & \multicolumn{1}{c}{other} & clean & \multicolumn{1}{c}{other} & clean & \multicolumn{1}{c}{other} & clean & \multicolumn{1}{c}{other}  \\ \midrule
&CNN$_{DR}$ & 3.61 & 11.37 & 3.77 & 11.08 & 3.04 & 9.52 & 3.69 & 9.38\\
 \cmidrule(l){2-10} 
&RNN$_{DS}$ & 7.64 & 22.02 & 7.70 & 22.60 & 4.88 & 16.00 & 5.18 & 16.55\\
\hspace{3mm}(1) CNN$_{DR}$ $\rightarrow$ RNN$_{DS}$ &\hspace{3mm}+ Sequence-level KD \cite{takashima-et-al:scheme} & 7.69 & 22.17 & 8.01 & 22.91 & 4.91 & 15.93 & 5.33 & 16.69\\
&\hspace{3mm}+ Guided CTC training \cite{kurata-et-al:scheme} & 7.32 & 22.16 & 7.63 & 22.66 & 5.37 & 17.33 & 5.55 & 17.61\\
&\hspace{3mm}\textbf{+ Ours} & \textbf{6.64} & \textbf{21.16} & \textbf{6.97} & \textbf{21.06} & \textbf{4.60} & \textbf{15.60} & \textbf{4.98} & \textbf{16.27}\\
\midrule
&CNN$_{DR}$ & 3.61 & 11.37 & 3.77 & 11.08 & 3.04 & 9.52 & 3.69 & 9.38 \\
\cmidrule(l){2-10} 
&CNN$_{Mini}$  & 8.66 & 23.28 & 8.85 & 24.26 & 4.83 & 15.53 & 5.24 & 16.40\\
\hspace{3mm}(2) CNN$_{DR}$ $\rightarrow$ CNN$_{Mini}$ &\hspace{3mm}+ Sequence-level KD \cite{takashima-et-al:scheme} & 8.96 & 23.73 & 9.10 & 24.81  & 5.16 & 15.54 & 5.48 & 16.91\\
&\hspace{3mm}+ Guided CTC training \cite{kurata-et-al:scheme} & 7.81 & 21.93 & 8.29 & 22.49 & 5.17 & 15.94 & 5.58 & 16.85 \\
&\hspace{3mm}\textbf{+ Ours} & \textbf{6.12} & \textbf{18.02} & \textbf{6.23} & \textbf{18.69} & \textbf{4.41} & \textbf{13.89} & \textbf{4.72} & \textbf{14.23}\\
\midrule
&RNN$_{DS1}$ & 6.64 & 21.16 & 6.97 & 21.06 & 4.60 & 15.60 & 4.98 & 16.27\\
\cmidrule(l){2-10}
&CNN$_{Mini}$ & 8.66 & 23.28 & 8.85 & 24.26 & 4.83 & 15.53 & 5.24 & 16.40\\
\hspace{3mm}(3)-(a) RNN$_{DS1}$ $\rightarrow$ CNN$_{Mini}$ &\hspace{3mm}+ Sequence-level KD \cite{takashima-et-al:scheme} & 8.17 & 22.43 & 8.46 & 23.34 & 5.17 & 15.60 & 5.53 & 16.52\\
&\hspace{3mm}+ Guided CTC training \cite{kurata-et-al:scheme} & 8.01 & 21.94 & 8.11 & 22.53 & 5.25 & 15.91 & 5.63 & 16.47 \\
&\hspace{3mm}\textbf{+ Ours} & \textbf{6.26} & \textbf{18.01} & \textbf{6.33} & \textbf{18.37} & \textbf{4.50} & \textbf{13.78} & \textbf{4.75} & \textbf{14.00} \\
\midrule 
&RNN$_{DS2}$ & 7.21 & 21.78 & 7.47 & 22.15 & 4.83 & 15.59 & 5.08 & 16.52\\
\cmidrule(l){2-10} 
&CNN$_{Mini}$ & 8.66 & 23.28 & 8.85 & 24.26 & 4.83 & 15.53 & 5.24 & 16.40\\
\hspace{3mm}(3)-(b) RNN$_{DS2}$ $\rightarrow$ CNN$_{Mini}$ &\hspace{3mm}+ Sequence-level KD \cite{takashima-et-al:scheme} & 7.49 & 20.81 & 7.55 & 21.84 & 5.08 & 15.30 & 5.22 &16.08\\
&\hspace{3mm}+ Guided CTC training \cite{kurata-et-al:scheme} & 7.74 & 21.29 & 7.85 & 21.88 & 5.26 & 15.61 & 5.69 & 16.24\\
&\hspace{3mm}\textbf{+ Ours} & \textbf{6.25} & \textbf{18.10} & \textbf{6.27} & \textbf{18.82} & \textbf{4.40} & \textbf{13.88} & \textbf{4.83} & \textbf{14.44}\\
\midrule
&RNN$_{DS1}$ & 6.64 & 21.16 & 6.97 & 21.06 & 4.60 & 15.60 & 4.98 & 16.27\\
\cmidrule(l){2-10}
&RNN$_{DS}$ & 7.64 & 22.02 & 7.70 & 22.60 & 4.88 & 16.00 & 5.18 & 16.55\\
\hspace{3mm}(4)-(a) RNN$_{DS1}$ $\rightarrow$ RNN$_{DS}$ &\hspace{3mm}+ Sequence-level KD \cite{takashima-et-al:scheme} & 7.41 & 21.51 & 7.58 & 22.23  & 5.02 & 16.29 & 5.23 & 17.16\\
&\hspace{3mm}+ Guided CTC training \cite{kurata-et-al:scheme} & 7.34 & 21.84 & 7.54 & 22.33 & 5.22 & 16.64 & 5.49 & 16.98\\
&\hspace{3mm}\textbf{+ Ours} & \textbf{6.51} & \textbf{20.28} & \textbf{6.77} & \textbf{20.58} & \textbf{4.63} & \textbf{15.37} & \textbf{4.95} & \textbf{15.66}\\
\midrule
&RNN$_{DS2}$ & 7.21 & 21.78 & 7.47 & 22.15 & 4.83 & 15.59 & 5.08 & 16.52\\
\cmidrule(l){2-10}
&RNN$_{DS}$ & 7.64 & 22.02 & 7.70 & 22.60 & 4.88 & 16.00 & 5.18 & 16.55 \\
\hspace{3mm}(4)-(b) RNN$_{DS2}$ $\rightarrow$ RNN$_{DS}$ &\hspace{3mm}+ Sequence-level KD \cite{takashima-et-al:scheme} & 7.39 & 21.76 & 7.56 & 22.26 & 5.05 & 16.41 & 5.30 & 17.33\\
&\hspace{3mm}+ Guided CTC training \cite{kurata-et-al:scheme} & 7.39 & 21.83 & 7.49 & 22.24 & 5.17 & 16.80 & 5.38 & 17.37\\
&\hspace{3mm}\textbf{+ Ours} & \textbf{6.52} & \textbf{19.98} & \textbf{6.80} & \textbf{20.93} & \textbf{4.54} & \textbf{15.14} & \textbf{4.90} & \textbf{16.21}\\
\bottomrule
\end{tabular}
\end{table*}

\begin{table}[t]
\centering
\caption{RERR (\%) on LibriSpeech dev-clean when greedy decoding was applied. We evaluated the performance at each stage of TutorNet in the case of CNN$_{DR}$ $\rightarrow$ RNN$_{DS}$.}
\label{cnnrnnrerr}
\begin{tabular}{lcc}
\toprule
\hspace{8mm}TutorNet & WER (\%) & RERR (\%) \\
\midrule
SKD  & 7.21 & 5.63\\
SKD+RKD w/o $M_{FW}$  & 6.74 & 11.78\\
SKD+RKD w/ $M_{FW}$ & 6.64 & 13.09\\
\bottomrule
\end{tabular}
\end{table}

We first experimented with TutorNet in CNN$_{DR}$ $\rightarrow$ RNN$_{DS}$ scenario. Table \ref{totalwer}-(1) summarizes the WER results on LibriSpeech, comparing the performance of the proposed method with other previous KD approaches.
The results show that most of the conventional KD methods did not always perform better than the student baseline, and in some cases, their performances were even worse.
In contrast, TutorNet always improved the performance of the original student.
As shown in Table \ref{cnnrnnrerr}, each stage of TutorNet was found to be useful for training the student. We observed that RNN$_{DS}$ can be effectively trained using the hidden representation of CNN$_{DR}$ via TutorNet.
Also, RKD with frame weighting contributed to improving the WER performance compared with the unweighted RKD.
Our best performance was achieved in case of using both RKD and SKD, which yielded WER 6.64\% (RERR 13.09 \%) with greedy decoding and WER 4.60 \% (RERR 5.74 \%) with beam-search decoding in the dev-clean dataset.

\begin{table}[t]
\centering
\caption{RERR (\%) on LibriSpeech dev-clean when greedy decoding was applied. We evaluated the performance at each stage of TutorNet when CNN$_{DR}$ $\rightarrow$ CNN$_{Mini}$.}
\label{cnncnnrerr}
\begin{tabular}{lcc}
\toprule
\hspace{8mm}TutorNet & WER (\%) & RERR (\%) \\
\midrule
SKD &  7.64 & 11.78\\
SKD+RKD w/o $M_{FW}$ & 6.40 & 26.10\\
SKD+RKD w/ $M_{FW}$ & 6.12 & 29.33\\
\bottomrule
\end{tabular}
\end{table}

\begin{table}[t]
\centering
\caption{RERR (\%) on LibriSpeech dev-clean when greedy decoding was applied. We evaluated the performance at each stage of TutorNet in the case of RNN$_{DS}$ $\rightarrow$ CNN$_{Mini}$.}
\label{rnncnnrerr}
\resizebox{\columnwidth}{!}{%
\begin{tabular}{clcc}
\toprule
Teacher model &\hspace{8mm}TutorNet & WER (\%) & RERR (\%) \\
\midrule
&SKD & 7.32 & 15.47\\
RNN$_{DS1}$&SKD+RKD w/o $M_{FW}$ & 6.40 & 26.10 \\
&SKD+RKD w/ $M_{FW}$ & 6.26 & 27.71\\
\midrule
&SKD & 6.65 & 23.21\\
RNN$_{DS2}$&SKD+RKD w/o $M_{FW}$ &  6.32 & 27.02\\
&SKD+RKD w/ $M_{FW}$ & 6.25 & 27.83\\
\bottomrule
\end{tabular}}
\end{table}

\subsubsection{CNN$_{DR}$ $\rightarrow$ CNN$_{Mini}$}
We also conducted experiments on KD between CNN$_{DR}$ and CNN$_{Mini}$. The main structural difference between the two models is that CNN$_{Mini}$ consists of depthwise separable convolutions with no dense residual connection. The results in Table \ref{totalwer}-(2) report that TutorNet showed significant improvements even when the teacher and student models had different convolution types. In the case of greedy decoding, compared to the previous results in Table \ref{totalwer}-(1), the guided CTC training \cite{kurata-et-al:scheme} achieved considerable improvement over the student baseline with WER 7.81\% for dev-clean.
Still, the proposed method showed the best performance in all configurations with WER 6.12 \% on the dev-clean dataset.
When applying beam-search decoding with LM, TutorNet achieved WER 4.41 \% on dev-clean, even though most of the conventional methods did not show the improvement over the baseline in CNN$_{DR}$ $\rightarrow$ CNN$_{Mini}$.
From Table \ref{cnncnnrerr}, we can also confirm that SKD and RKD significantly improved the WER performance of the original student in all cases, and the best performance was obtained when both were applied.

\subsubsection{RNN$_{DS}$ $\rightarrow$ CNN$_{Mini}$}

In addition to the previous experiments, we proceeded to verify whether CNN$_{Mini}$ can benefit from the guidance of RNN$_{DS}$. Since our distilled RNN$_{DS}$ in Section \rom{5}-A-1 (Table \ref{totalwer}-(1)) performed better than the original DeepSpeech2 in the OpenSeq2Seq toolkit, we could employ it as the teacher in this scenario.
In this experiment, we applied two different RNN$_{DS}$ models as the teachers, where both had previously been distilled in CNN$_{DR}$ $\rightarrow$ RNN$_{DS}$ scenario.
Therefore, RNN$_{DS}$ $\rightarrow$ CNN$_{Mini}$ can be rewritten as (CNN$_{DR}$ $\rightarrow$ RNN$_{DS}$) $\rightarrow$ CNN$_{Mini}$.
To distinguish the two RNN-based teachers, we used numbers, i.e., RNN$_{DS1}$ and RNN$_{DS2}$.
\begin{figure}[t]
	{
		\begin{center}
			\begin{tabular}{c}
				\includegraphics[height=5cm]{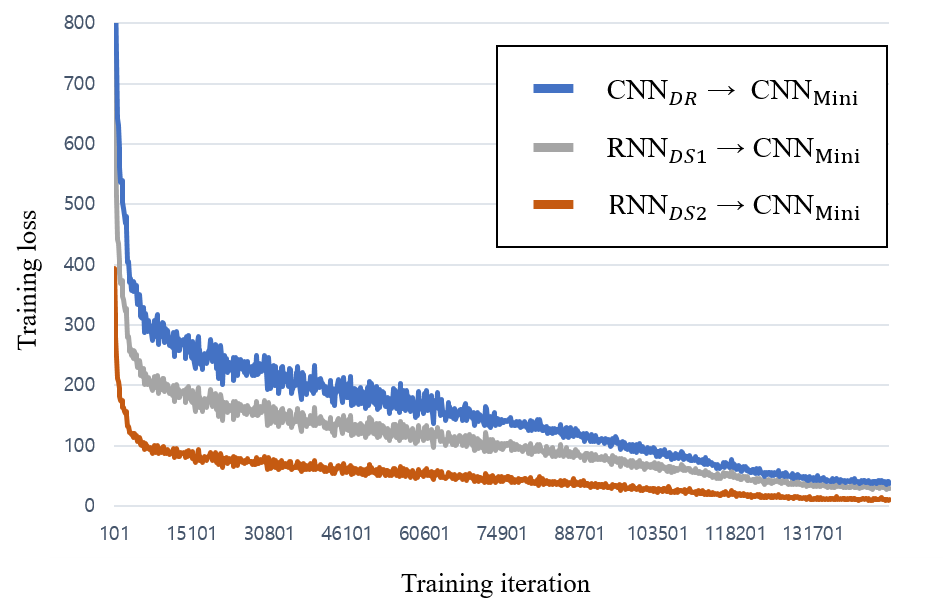}
			\end{tabular}
		\end{center}
	}
	\caption{Analysis of the training loss $\mathcal{L}_{CTC-SKD}$. We compared  CNN$_{Mini}$ distilled from CNN$_{DR}$, RNN$_{DS1}$, and RNN$_{DS2}$, respectively.}
	\label{cnnskd}
\end{figure}

First, we applied RNN$_{DS1}$ (WER 6.64 \% with greedy decoding) as the teacher. WERs on LibriSpeech corpus are shown in Table \ref{totalwer}-(3)-(a). 
From the results, we observed that TutorNet helped CNN$_{Mini}$ to benefit from the RNN-based teacher, while achieving better performance than the other conventional methods.
Interestingly, when both SKD and RKD were applied, the distilled CNN$_{Mini}$ performed better than its teacher in all cases, including greedy decoding and beam-search decoding. In the dev-clean dataset, with greedy decoding, TutorNet achieved WER 6.26 \%, although that of RNN$_{DS1}$ was 6.64 \%. In the case of beam-search decoding, TutorNet also outperformed the teacher, as it achieved WER 4.50 \% on dev-clean dataset. From Table \ref{rnncnnrerr}, we can discover some interesting results of SKD training. When applying SKD, RNN$_{DS1}$ $\rightarrow$ CNN$_{Mini}$ provided WER 7.32 \% with greedy decoding. On the other hand, as presented in Table \ref{cnncnnrerr}, CNN$_{DR}$ $\rightarrow$ CNN$_{Mini}$ showed WER 7.64 \%. In terms of the WER performance, though RNN$_{DS1}$ (WER 6.64 \% with greedy decoding) was 3.03 \%p \footnote{We used the term ``\%p" as a percentage point, which means the arithmetic difference between two percentages.} worse than CNN$_{DR}$ (WER 3.61 \% with greedy decoding), the performance of the distilled student was 0.32 \%p better. Considering that CNN$_{DR}$ teacher required high computational resources (400 epochs with 8 Tesla V100 GPUs, each with 32GB of memory) to achieve WER 3.61 \%, RNN$_{DS}$ (50 epochs with 3 Titan V GPUs, each with 16GB of memory) could be an efficient alternative for transferring softmax-level knowledge.

From these experimental results regarding SKD, we verified that the teacher with the high WER performance did not necessarily help the student model's training. In some particular cases, the poor teacher could be more supportive for training the student. To further check the effect of the teacher's performance on SKD training, we adopted RNN$_{DS2}$ (WER 7.21 \% with greedy decoding on dev-clean) as another RNN-based teacher model, where the knowledge had been transferred from CNN$_{DR}$ in Section \rom{5}-A-1 (Table \ref{cnnrnnrerr}). Table \ref{totalwer}-(3)-(b) gives the WER results on LibriSpeech corpus. Compared to the competing KD methods, TutorNet showed better WER improvements in all cases.
Also, TutorNet allowed CNN$_{Mini}$ to outperform its teacher when applying both SKD and RKD, provided with WER 6.25 \% with greedy and WER 4.40 \% with beam-search decoding.
As given in Table \ref{rnncnnrerr}, RNN$_{DS2}$ $\rightarrow$ CNN$_{Mini}$ achieved WER 6.65 \% with SKD training, though RNN$_{DS1}$ $\rightarrow$ CNN$_{Mini}$ provided WER 7.32 \%.
This means that the softmax prediction of RNN$_{DS2}$ (WER 7.21 \% with greedy decoding) was considered more informative than that of RNN$_{DS1}$ (WER 6.64 \% with greedy decoding).
When we applied RKD and SKD altogether, the performance of RNN$_{DS2}$ $\rightarrow$ CNN$_{Mini}$ was slightly better than that of  RNN$_{DS1}$ $\rightarrow$ CNN$_{Mini}$, but the difference was negligible.

Next, we compared ${L}_{CTC-SKD}$ when transferring from different teachers to the same CNN$_{Mini}$. Fig. \ref{cnnskd} displays the variation of the training loss over time with three different cases: (1) CNN$_{DR}$ $\rightarrow$ CNN$_{Mini}$, (2) RNN$_{DS1}$ $\rightarrow$ CNN$_{Mini}$, and (3) RNN$_{DS2}$ $\rightarrow$ CNN$_{Mini}$. With SKD training, RNN$_{DS2}$ $\rightarrow$ CNN$_{Mini}$ showed faster convergence than the others, which indicates that the teacher model with the highest WER performance did not necessarily help the student model's training.
In other words, RNN$_{DS2}$ can be more supportive in distilling knowledge, notwithstanding its smaller parameter size (13.1 M parameters) and worse performance (WER 7.21 \% with greedy decoding) compared to CNN$_{DR}$ (332.6 M parameters / WER 3.61 \% with greedy decoding). 
\begin{table*}[t]
\centering
\caption{Comparison of WER (\%) and RERR (\%) on LibriSpeech in the case of (CNN$_{DR}$ \& RNN$_{DS}$) $\rightarrow$ CNN$_{Mini}$. We evaluated the performance in the case of selecting suitable teachers for SKD and RKD. The best result is in bold.}
\label{mix}
\begin{tabular}{@{}cccccccccccc@{}}
\toprule
\multicolumn{1}{c}{} &  &  & & \multicolumn{4}{c}{WER (\%)} & \multicolumn{4}{c}{RERR (\%)} \\ \cmidrule(l){5-12} 
\multicolumn{1}{c}{Student model} & SKD teacher & RKD teacher & LM & \multicolumn{2}{c}{dev} & \multicolumn{2}{c}{test} & \multicolumn{2}{c}{dev} & \multicolumn{2}{c}{test} \\ \cmidrule(l){5-12}
\multicolumn{1}{c}{} &  & &  & clean & other & clean & other & \multicolumn{1}{c}{clean} & \multicolumn{1}{c}{other} & \multicolumn{1}{c}{clean} & \multicolumn{1}{c}{other} \\ \midrule
\multirow{4}{*}{CNN$_{Mini}$}& - & - & \multirow{4}{*}{-} & 8.66 & 23.28 & 8.85 & 24.26 & - & - & - & -\\ 
& RNN$_{DS2}$ & RNN$_{DS2}$ &  & 6.25 & 18.10 & 6.27 & 18.82 & 27.83 & 22.25 & 29.15 & 22.42 \\
& CNN$_{DR}$ & CNN$_{DR}$ &  & 6.12 & \textbf{18.02} & 6.23 & 18.69 & 29.33 & \textbf{22.59} & 29.60 & 22.96 \\
 & RNN$_{DS2}$ & CNN$_{DR}$ &  & \textbf{6.03} & \textbf{18.02} & \textbf{6.20} & \textbf{18.52} & \textbf{30.37} & \textbf{22.59} & \textbf{29.94} & \textbf{23.66} \\ \midrule
\multirow{4}{*}{CNN$_{Mini}$} & - & - & \multirow{4}{*}{4-gram} & 4.83 & 15.53 & 5.24 & 16.40 & - & - & - & -
\\
& RNN$_{DS2}$ & RNN$_{DS2}$ & & \textbf{4.40} & 13.88 & 4.83 & 14.44 & \textbf{8.90} & 10.62 & 7.82 & 11.95 \\
& CNN$_{DR}$ & CNN$_{DR}$ &  & 4.41 & 13.89 & 4.72 & \textbf{14.23} & 8.70 & 10.56 & 9.92 & \textbf{13.23} \\
 & RNN$_{DS2}$ & CNN$_{DR}$ &  & \textbf{4.40} & \textbf{13.78} & \textbf{4.62} & 14.32 & \textbf{8.90} & \textbf{11.27} & \textbf{11.83} & 12.68\\
\bottomrule
\end{tabular}
\end{table*}

\begin{table*}[t]
\centering
\caption{Comparison of WER (\%) and RERR (\%) on LibriSpeech in the case of (CNN$_{DR}$ \& RNN$_{DS}$) $\rightarrow$ QuartzNet. The best result is in bold.}
\label{quartznet}
\begin{tabular}{@{}ccccccccccccc@{}}
\toprule
\multicolumn{1}{c}{} &  &  & && \multicolumn{4}{c}{WER (\%)} & \multicolumn{4}{c}{RERR (\%)} \\ \cmidrule(l){6-13} 
\multicolumn{1}{c}{Student model} & SKD teacher & RKD teacher & LM & Augmentation & \multicolumn{2}{c}{dev} & \multicolumn{2}{c}{test} & \multicolumn{2}{c}{dev} & \multicolumn{2}{c}{test} \\ \cmidrule(l){6-13}
\multicolumn{1}{c}{} &  & &  & & clean & other & clean & other & \multicolumn{1}{c}{clean} & \multicolumn{1}{c}{other} & \multicolumn{1}{c}{clean} & \multicolumn{1}{c}{other} \\ \midrule
\multirow{2}{*}{QuartzNet}& - & - & \multirow{2}{*}{-} & \multirow{2}{*}{-} & 5.66 & 16.57 & 5.85 & 17.18 & - & - & - & -\\ 
 & RNN$_{DS2}$ & CNN$_{DR}$ &  &  & \textbf{4.95} & \textbf{15.08} & \textbf{4.95} & \textbf{15.46} & \textbf{12.54} & \textbf{8.99} & \textbf{15.38} & \textbf{10.01} \\
 \midrule
 \multirow{2}{*}{QuartzNet}& - & - & \multirow{2}{*}{-} & \multirow{2}{*}{SpecAugment \cite{specaug:scheme}} & 5.33 & 15.32 & 5.33 & 15.48 & -& -&-&-\\
 & RNN$_{DS2}$ & CNN$_{DR}$ &  & & \textbf{4.68} & \textbf{13.92} & \textbf{4.68} & \textbf{14.06} & \textbf{12.20} & \textbf{9.14} & \textbf{12.20} & \textbf{9.17} \\
 \midrule
 \multirow{2}{*}{QuartzNet}& - & - & \multirow{2}{*}{4-gram} & \multirow{2}{*}{-} & 4.03 & 12.36 & 4.66 & 13.25 & - & - & - & -\\ 
 & RNN$_{DS2}$ & CNN$_{DR}$ &  &  &\textbf{3.95} & \textbf{12.25} & \textbf{4.32} & \textbf{12.56} & \textbf{1.99} & \textbf{0.89} & \textbf{7.30} & \textbf{5.21} \\
 \midrule
  \multirow{2}{*}{QuartzNet}& - & - & \multirow{2}{*}{4-gram} & \multirow{2}{*}{SpecAugment \cite{specaug:scheme}} & 4.01 & 11.44 & 4.16 & 11.50 & - & - & - & -\\ 
 & RNN$_{DS2}$ & CNN$_{DR}$ &  &  & \textbf{3.54} & \textbf{10.78} & \textbf{3.86} & \textbf{11.14} & \textbf{11.72} & \textbf{5.77} & \textbf{7.21}& \textbf{3.13}   \\
\bottomrule
\end{tabular}
\end{table*}
\subsubsection{RNN$_{DS}$ $\rightarrow$ RNN$_{DS}$}
\begin{figure}[t]
	{
		\begin{center}
			\begin{tabular}{c}
				\includegraphics[height=5cm]{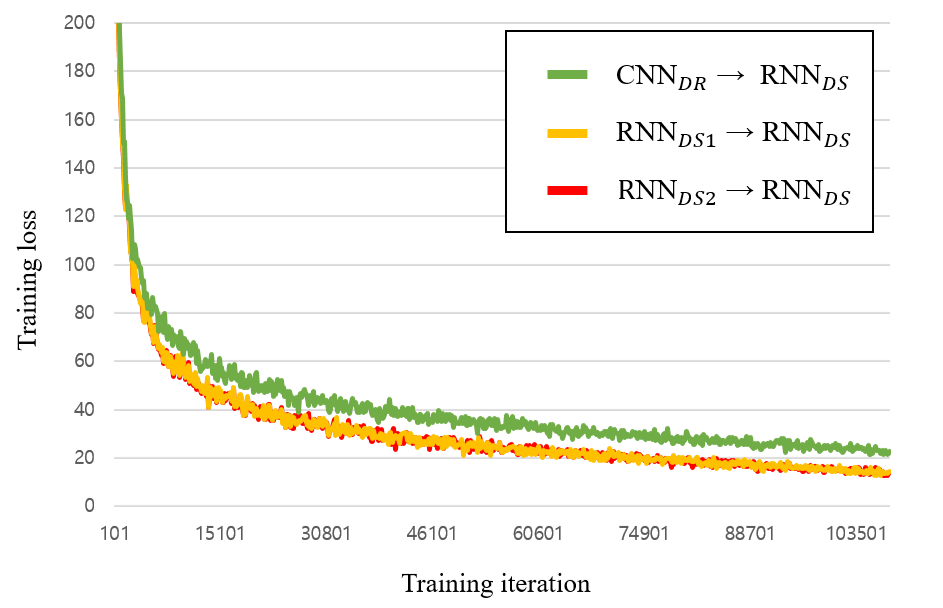}
			\end{tabular}
		\end{center}
	}
	\caption{Analysis of the training loss $\mathcal{L}_{CTC-SKD}$. We compared  RNN$_{DS}$ distilled from CNN$_{DR}$, RNN$_{DS1}$ and RNN$_{DS2}$, respectively.}
	\label{rnnskd}
\end{figure}
In the previous experiments covered in Section \rom{5}-A-1, \rom{5}-A-2, and \rom{5}-A-3, we mainly paid attention to how well TutorNet can distill the knowledge between networks with different topologies. 
On top of that, we tried to verify that TutorNet still works well in RNN$_{DS}$ $\rightarrow$ RNN$_{DS}$ transfer, which is a typical KD case. 
To maintain the same model configuration, we repetitively used RNN$_{DS1}$ (WER 6.64 \% with greedy decoding) and RNN$_{DS2}$ (WER 7.21 \% with greedy decoding) as the teachers. Since both RNN$_{DS1}$ and RNN$_{DS2}$ had previously been transferred from CNN$_{DR}$ in Section \rom{5}-A, RNN$_{DS}$ $\rightarrow$ RNN$_{DS}$ can be rewritten as (CNN$_{DR}$ $\rightarrow$ RNN$_{DS}$) $\rightarrow$ RNN$_{DS}$.

From Table \ref{totalwer}-(4)-(a), we can confirm that TutorNet gave better WER improvements than the other conventional approaches. It means that TutorNet is well applied in the case of normal KD, where both teacher and student models had the same model architecture.
Our best results were achieved when training the student with both RKD and SKD. 
Compared to the previous results in Table \ref{totalwer}-(1), RNN$_{DS1}$ $\rightarrow$ RNN$_{DS}$ performed better than CNN$_{DR}$ $\rightarrow$ RNN$_{DS}$, except for dev-clean dataset with beam-search decoding. This means that RNN$_{DS1}$ could be considered more supportive than CNN$_{DR}$ in transferring the knowledge to RNN$_{DS}$. When we applied both RKD and SKD, RNN$_{DS1}$ $\rightarrow$ RNN$_{DS}$ provided WER 6.51 \% in the dev-clean dataset with greedy decoding, which was 0.13 \%p better than that of CNN$_{DR}$ $\rightarrow$ RNN$_{DS}$.

Table \ref{totalwer}-(4)-(b) shows the WER results obtained from the RNN$_{DS2}$ $\rightarrow$ RNN$_{DS}$ scenario.
In the case of beam-search decoding, RNN$_{DS2}$ $\rightarrow$ RNN$_{DS}$ performed better than RNN$_{DS1}$ $\rightarrow$ RNN$_{DS}$, except for test-clean dataset. The results show that most of the conventional KD methods did not always perform better than the student baseline, and in some cases, their performances were even worse.
In contrast, TutorNet always improved the performance of the original student in both RNN$_{DS1}$ $\rightarrow$ RNN$_{DS}$ and RNN$_{DS2}$ $\rightarrow$ RNN$_{DS}$.

In addition, we compared $\mathcal{L}_{CTC-SKD}$ when distilling from different teachers to the same RNN$_{DS}$. In Fig. \ref{rnnskd}, We plot the change in $\mathcal{L}_{CTC-SKD}$ over time in three different cases: (1) CNN$_{DR}$ $\rightarrow$ RNN$_{DS}$, (2) RNN$_{DS1}$ $\rightarrow$ RNN$_{DS}$, and (3) RNN$_{DS2}$ $\rightarrow$ RNN$_{DS}$. In the case of training SKD, RNN$_{DS}$ $\rightarrow$ RNN$_{DS}$ scenarios showed faster convergence than CNN$_{DR}$ $\rightarrow$ RNN$_{DS}$. Considering the high performance of CNN$_{DR}$ (WER 3.61 \% with greedy decoding), it is interesting that RNN$_{DS1}$ $\rightarrow$ RNN$_{DS}$ and RNN$_{DS2}$ $\rightarrow$ RNN$_{DS}$ showed faster convergence than CNN$_{DR}$ $\rightarrow$ RNN$_{DS}$. Unlike Fig. \ref{cnnskd}, there was no significant difference between RNN$_{DS1}$ $\rightarrow$ RNN$_{DS}$ and RNN$_{DS2}$ $\rightarrow$ RNN$_{DS}$.

\subsubsection{(CNN$_{DR}$ \&  RNN$_{DS}$) $\rightarrow$ CNN$_{Mini}$}

\begin{table*}[t]
\centering
\caption{Comparison of CER (\%) on AISHELL-2. All results were evaluated based only on greedy decoding. ~``Ours" denotes TutorNet with SKD and RKD. The best result is in bold.}
\label{aishell}
\begin{tabular}{@{}clcccccc@{}}
\toprule
& & \multicolumn{6}{c}{CER (\%)}  \\ \cmidrule(l){3-8} 
KD & \hspace{15mm}Model & \multicolumn{3}{c}{dev} & \multicolumn{3}{c}{test}  \\ \cmidrule(l){3-8} 
& & iOS & \multicolumn{1}{c}{Android} & Mic & iOS & \multicolumn{1}{c}{Android} & Mic  \\ \midrule
\multirow{4}{*}{\hspace{3mm}CNN$_{DR}$ $\rightarrow$ RNN$_{DS}$}&CNN$_{DR}$ & 9.69 & 11.48 & 12.23 & 9.37 & 10.84 & 11.84\\
 \cmidrule(l){2-8} 
&RNN$_{DS}$ & 13.40 & 16.66 & 17.93 & 12.51 & 14.08 & 16.95\\
&\hspace{3mm}+ Guided CTC training \cite{kurata-et-al:scheme} & 12.74 & 14.98 & 15.55 & 12.36 & 13.91 & 14.80\\
&\hspace{3mm}\textbf{+ Ours} & \textbf{12.30} & \textbf{14.22} & \textbf{14.77} & \textbf{12.07}& \textbf{13.52} & \textbf{14.35}\\
\midrule
\multirow{4}{*}{\hspace{3mm}CNN$_{DR}$ $\rightarrow$ CNN$_{Mini}$} & CNN$_{DR}$ & 9.69 & 11.48 & 12.23 & 9.37 & 10.84 & 11.84\\
\cmidrule(l){2-8} 
&CNN$_{Mini}$  & 11.77 & 14.23 & 15.03 & 11.38 & 12.71 & 14.27\\
&\hspace{3mm}+ Guided CTC training \cite{kurata-et-al:scheme} & 11.34 & 13.33 & 14.49 & 10.87 &12.40& 13.88\\
&\hspace{3mm}\textbf{+ Ours} & \textbf{10.65} & \textbf{12.74} & \textbf{13.99} & \textbf{10.08} &\textbf{11.41}&\textbf{13.38}\\
\bottomrule
\end{tabular}
\end{table*}
The results of the previous experiments, especially in Sections \rom{5}-A-2 and \rom{5}-A-3, suggest that the teacher with the highest WER performance did not necessarily help the training of the student, and the best-performing teacher was different for each level of knowledge. For instance, even though RNN$_{DS2}$ had worse achievement than the other teachers, it was shown to be more supportive in transferring the softmax-level knowledge. Meanwhile, CNN$_{DR}$ was more effective in distilling the representation-level knowledge to CNN$_{Mini}$. These observations lead us to an interesting perspective: Can we get more improved results if we select separate teachers for RKD and SKD? 

In order to verify this question, we conducted additional experiments applying CNN$_{DR}$ as an RKD teacher and RNN$_{DS2}$ as an SKD teacher. CNN$_{Mini}$ was adopted as the target of distillation.
The WER results are described in Table \ref{mix}. CNN$_{Mini}$ distilled from suitable teachers for each stage showed better performance. The performance difference between CNN$_{DR}$ $\rightarrow$ CNN$_{Mini}$ and (CNN$_{DR}$ \&  RNN$_{DS}$) $\rightarrow$ CNN$_{Mini}$ looks a little negligible, but (CNN$_{DR}$ \&  RNN$_{DS}$) $\rightarrow$ CNN$_{Mini}$ showed better WER improvements in all cases with greedy decoding. Also, when applying beam-search decoding, (CNN$_{DR}$ \&  RNN$_{DS}$) $\rightarrow$ CNN$_{Mini}$ performed better than the others in most cases, including dev-clean, dev-other, and test-clean dataset. From these results, we believe that the proposed method enables much more flexible model selection in KD, which means that we can select suitable teacher models for each stage regardless of the model type via TutorNet. Such flexibility in KD can give more possibilities for better achievement.

\subsubsection{(CNN$_{DR}$ \&  RNN$_{DS}$) $\rightarrow$ QuartzNet}
In Section \rom{5}-A-5, we adopted CNN$_{DR}$ as an RKD teacher and RNN$_{DS2}$ as an SKD teacher, which gave a better performance in our experiments. Based on these results, we tried to apply the same teacher selection (RKD teacher: CNN$_{DR}$, SKD teacher: RNN$_{DS2}$) to QuartzNet, since QuartzNet can achieve near state-of-the-art accuracy among CTC models while having fewer parameters. Also, in order to further improve the performance, we applied SpecAugment \cite{specaug:scheme} as an augmentation technique. Table  \ref{quartznet} gives the WER results on LibriSpeech. With greedy decoding, TutorNet (w/o SpecAugment) showed WER 4.95 \% and RERR 12.54 \% on the dev-clean dataset. Our best performance was achieved when applying TutorNet with SpecAugment, which yielded WER 4.68 \% and RERR 12.20 \%. In the case of beam-search decoding, (CNN$_{DR}$ \&  RNN$_{DS}$) $\rightarrow$ QuartzNet (w/ SpecAugment) provided 3.54 \% on dev-clean, which was the highest performance in our experiments.
\subsection{AISHELL-2}

In order to show the versatility of the proposed method, we also conducted our experiment on AISHELL-2:
\begin{itemize}
    \item CNN$_{DR}$ $\rightarrow$ RNN$_{DS}$: From CNN-based model (Jasper DR) to RNN-based model (DeepSpeech2)
    \item CNN$_{DR}$ $\rightarrow$ CNN$_{Mini}$: From CNN-based model (Jasper DR) to CNN-based model (Jasper Mini)
\end{itemize}
We experimented with TutorNet in both CNN$_{DR}$ $\rightarrow$ RNN$_{DS}$ and CNN$_{DR}$ $\rightarrow$ CNN$_{Mini}$ scenarios with AISHELL-2 dataset. For the teacher model, we used pre-trained Jasper DR (CER \% on iOS dev dataset), which was provided by NeMo toolkit. When training the CNN-based and RNN-based student models, we utilized OpenSeq2Seq toolkit. Table \ref{aishell} summarizes the results on AISHELL-2. The results show that TutorNet still works well with the Mandarin dataset. When distilling knowledge to RNN$_{DS}$, the student provided 12.30 \% in dev iOS dataset.
In the case of CNN$_{DR}$ $\rightarrow$ CNN$_{Mini}$, TutorNet achieved WER 10.65 \% and RERR 9.52 \% in dev iOS dataset.

\subsection{The Applicability of TutorNet to the Other End-to-End Speech Recognition Models}

In the previous experiments, we mainly focus on KD between CTC-based models, especially across different neural networks. For speech recognition, besides the CTC-based model, there have been various types of end-to-end models such as Transformer and AED model. If we can transfer the knowledge regardless of the types of models, a more flexible KD will be possible. To check the applicability of the proposed method to the other types of models, we conducted two different scenarios with LibriSpeech dataset:

\begin{itemize}
    \item Transformer $\rightarrow$ CTC-based model: From Transformer to CTC-based model (DeepSpeech2)
    \item Transformer $\rightarrow$ AED model: From Transformer to AED model (VGG-BLSTM)
\end{itemize}

\subsubsection{Transformer $\rightarrow$ CTC-based model}
\begin{figure*}[t]
	\begin{center}
			\begin{tabular}{c}
				\includegraphics[height=5.5cm]{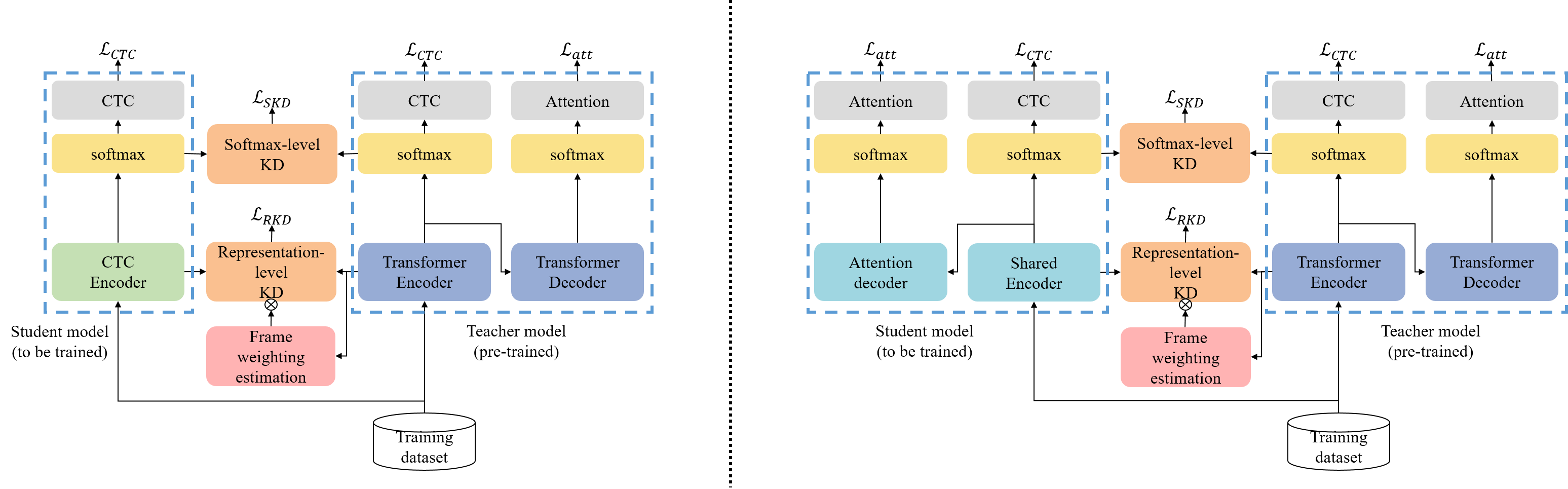}
			\end{tabular}
		\end{center}
	\caption{The procedure of TutorNet when transferring knowledge from Transformer. The left figure shows KD from Transformer to CTC-based model, and the right figure illustrates KD from Trnasformer to attention-based encoder-decoder (AED) model.}
	\label{trans}
\end{figure*}
\begin{table}[t]
\centering
\caption{WER (\%) on LibriSpeech in the case of Transformer $\rightarrow$ CTC model with greedy decoding. The best result is in bold. DNC denotes ‘Did not converge’.}
\label{transformerctc}
\begin{tabular}{@{}cccccc@{}}
\toprule
\multicolumn{1}{c}{} & &  \multicolumn{4}{c}{WER (\%)} \\ \cmidrule(l){3-6} 
\multicolumn{1}{c}{Model} &  TutorNet & \multicolumn{2}{c}{dev} & \multicolumn{2}{c}{test} \\ \cmidrule(l){3-6}
\multicolumn{1}{c}{} &  &  clean & other & clean & other \\ \midrule
Transformer& - & 3.21 & 8.58 & 3.45 & 8.45  \\ 
\midrule
\multirow{4}{*}{CTC-based model}& -  & \multicolumn{4}{c}{DNC} \\
& RKD for 100 steps & 8.82 & 22.72 & 8.84 & 23.55\\
& RKD & 8.26 & 22.05 & 8.20 & 22.73 \\
& RKD+SKD & \textbf{7.86} & \textbf{21.95} & \textbf{8.11} & \textbf{22.62}\\ 
\bottomrule
\end{tabular}
\end{table}
\begin{table*}[t]
\centering
\caption{Comparison of WER (\%) and RERR (\%) on LibriSpeech in the case of Transformer $\rightarrow$ AED model. The best result is in bold.}
\label{transformeraed}
\begin{tabular}{@{}ccccccccccc@{}}
\toprule
\multicolumn{1}{c}{} &  & & \multicolumn{4}{c}{WER (\%)} & \multicolumn{4}{c}{RERR (\%)} \\ \cmidrule(l){4-11} 
\multicolumn{1}{c}{Model} & TutorNet & LM & \multicolumn{2}{c}{dev} & \multicolumn{2}{c}{test} & \multicolumn{2}{c}{dev} & \multicolumn{2}{c}{test} \\ \cmidrule(l){4-11}
\multicolumn{1}{c}{} &  &  & clean & other & clean & other & \multicolumn{1}{c}{clean} & \multicolumn{1}{c}{other} & \multicolumn{1}{c}{clean} & \multicolumn{1}{c}{other} \\ \midrule
Transformer&  - & - & 3.21 & 8.58 & 3.45 & 8.45 & - & - & - & -\\ 
\midrule
\multirow{3}{*}{AED}& - & \multirow{3}{*}{-} & 5.80 & 16.80 & 5.75 & 17.50 & -  & - & - & -\\
& RKD &  & 4.89  & 15.15 & 4.91 & 15.65 & 15.69 & 9.82 & 14.61 & 10.57\\
 & RKD+SKD &  & \textbf{4.66} & \textbf{15.00} & \textbf{4.87} & \textbf{15.35} & \textbf{19.67} & \textbf{10.71} & \textbf{15.30} &  \textbf{12.29} \\ \midrule
Transformer& - & RNNLM & 2.82 & 7.21 & 3.07 & 7.17 & - & - & - & -\\ 
\midrule
\multirow{3}{*}{AED}& - & \multirow{3}{*}{RNNLM} & 4.00 & 12.49 & 4.25 & 13.04 & - & - & - & - \\
& RKD &  & 3.85 & 11.66 & 3.86 & 12.38 & 3.75 & 6.65 & 9.18 & 5.06 \\
 & RKD+SKD &  & \textbf{3.57} & \textbf{11.16} & \textbf{3.68} & \textbf{11.80} & \textbf{10.75} & \textbf{10.65} & \textbf{13.41} & \textbf{9.51} \\
\bottomrule
\end{tabular}
\end{table*}

Firstly, we conducted experiments on KD from Transformer model to the CTC-based model (RNN$_{DS}$). In the case of Transformer, we used the pre-trained model provided by the ESPnet toolkit. When training the CTC-based student, we utilized the OpenSeq2Seq toolkit. Since Transformer commonly uses BPE tokens as the output units, the CTC-based model was trained with the same BPE units of Transformer (about 5000 tokens). 
The detailed procedure for Transformer $\rightarrow$ CTC-based model is illustrated in Fig. \ref{trans}. For RKD, we selected the last layer of the Transformer encoder to transfer the hidden representation. After the RKD initialization, via SKD, the CTC-based model was trained with the softmax values provided by Transformer.
However, there was a convergence problem in training the CTC-based baseline with BPE units.
When the CTC-based model was trained with a single BPE CTC loss, random initialization and without any pre-training, it failed to converge, as in \cite{garg-et-al:scheme, audhkhasi-et-al:scheme}.
Therefore, we adopted the CTC model, initialized with RKD over the first 100 steps, as the student baseline for the comparison of the performance. Table \ref{transformerctc} summarizes the WER results on LibriSpeech. From the experimental results, we verified that the proposed method not only helps the convergence of the BPE-level CTC model but also can connect Transformer/CTC models in the KD task. When appying both RKD and SKD, the BPE-level CTC model achieved WER 7.86 \% in dev clean.

\subsubsection{Transformer $\rightarrow$ AED model}

In addition to the previous experiments of Transformer $\rightarrow$ CTC-based model, we proceeded to verify whether the AED model can benefit from the guidance of Transformer model via TutorNet. We employed a pre-trained Transformer model, provided by ESPnet toolkit, as the teacher model. We also trained the AED student model with the ESPnet toolkit. The detailed procedure of the proposed method is shown in Fig. \ref{trans}. For RKD, we selected the encoder's last layer to transfer the hidden representation for both teacher and student models. When training SKD, considering that both models are based on hybrid CTC/attention, we applied the softmax value on the CTC side. Table \ref{transformeraed} reports the WER and RERR performance of each stage. From the results, it is confirmed that TutorNet showed significant improvements in Transformer $\rightarrow$ AED model. Each stage of TutorNet was found to be useful for training the AED student model, which means that TutorNet is well applied in the case of Transformer $\rightarrow$ AED model.
In the dev-clean dataset, the student with RKD achieved WER 4.89 \% with greedy decoding and WER 3.85 \% during RNNLM decoding. It means that the AED model can be effectively trained using the hidden representation of Transformer via TutorNet. Our best performance of AED model was achieved in the case of using both RKD and SKD, provided with WER 4.66 \% with greedy and WER 3.57 \% with RNNLM. It is interesting that SKD does not only matter for the CTC loss but also help to improve the attention loss in the case of the hybrid CTC/attention model.

\section{Conclusion}
In this paper, we proposed a new KD method, TutorNet.
This framework copes with the shortcoming of the conventional KD, limiting the flexibility of model selection since the student model structure should be similar to that of the given teacher.
Through a number of experiments on LibriSpeech and AISHELL-2, we confirmed that TutorNet significantly contributes to improving the WER performance of the distilled student in an unexplored setting where the architecture of the student is inherently different from that of the teacher. In some particular configurations, it allows the student to outperform its teacher.
Furthermore, selecting a suitable teacher model for each training is possible via TutorNet, implying that we can have more flexibility in selecting teacher or student models.
For speech recognition, there have been various types of end-to-end models, and each model has its own advantages.
As various end-to-end speech recognition models can be flexibly selected, TutorNet represents a significant step toward KD in the speech recognition task.
We expect the application of TutorNet not to be restricted to the modality or model architecture of the tasks and would like to examine its utility via cross-domain studies in the future.


\ifCLASSOPTIONcaptionsoff
  \newpage
\fi

\bibliographystyle{unsrt}
\bibliography{Trans.bib}

\begin{thebibliography}{10}

\bibitem{graves-et-al:scheme}
A.~Graves, S.~Fern{\'a}ndez, F.~Gomez, and J.~Schmidhuber.
\newblock Connectionist temporal classification: labelling unsegmented sequence
  data with recurrent neural networks.
\newblock In {\em Proc. ICML}, pages 369--376, 2006.

\bibitem{chorowski-et-al:scheme}
J.~K. Chorowski, D.~Bahdanau, D.~Serdyuk, K.~Cho, and Y.~Bengio.
\newblock Attention-based models for speech recognition.
\newblock In {\em Proc. NIPS}, pages 577--585, 2015.

\bibitem{graves-rnn-t:scheme}
A.~Graves.
\newblock Sequence transduction with recurrent neural networks.
\newblock In {\em Proc. ICML Workshop on Representation Learning}, 2012.

\bibitem{amodei-et-al:scheme}
D.~Amodei, S.~Ananthanarayanan, R.~Anubhai, J.~Bai, E.~Battenberg, C.~Case,
  J.~Casper, B.~Catanzaro, Q.~Cheng, G.~Chen, et~al.
\newblock Deep speech 2: end-to-end speech recognition in english and mandarin.
\newblock In {\em Proc. ICML}, pages 173--182, 2016.

\bibitem{collobert-et-al:scheme}
R.~Collobert, C.~Puhrsch, and G.~Synnaeve.
\newblock Wav2letter: an end-to-end convnet-based speech recognition system.
\newblock {\em arXiv preprint arXiv:1609.03193}, 2016.

\bibitem{li-et-al:scheme}
J.~Li, V.~Lavrukhin, B.~Ginsburg, R.~Leary, O.~Kuchaiev, J.~M. Cohen,
  H.~Nguyen, and R.~T. Gadde.
\newblock Jasper: An end-to-end convolutional neural acoustic model.
\newblock {\em arXiv preprint arXiv:1904.03288}, 2019.

\bibitem{quartznet:scheme}
S.~Kriman, K.~Beliaev, B.~Ginsburg, J.~Huang, O.~Kuchaiev, V.~Lavrukhin,
  R.~Leary, J.~Li, and Y.~Zhang.
\newblock Quartznet: deep automatic speech recognition with 1d time-channel
  separable convolutions.
\newblock {\em arXiv preprint arXiv:1910.10261}, 2019.

\bibitem{transformer:scheme}
A.~Vaswani et~al.
\newblock Attention is all you need.
\newblock In {\em Proc. NIPS}, pages 5998–--6008, 2017.

\bibitem{hinton_kd-et-al:scheme}
G.~Hinton, O.~Vinyals, and J.~Dean.
\newblock Distilling the knowledge in a neural network.
\newblock In {\em Proc. NIPS Workshop Deep Learn.}, 2014.

\bibitem{dodeep:scheme}
J.~Ba and R.~Caruana.
\newblock Do deep nets really need to be deep?
\newblock In {\em Proc. NIPS}, pages 2654–--2662, 2014.

\bibitem{romero-et-al:scheme}
A.~Romero, N.~Ballas, S.~E. Kahou, A.~Chassang, C.~Gatta, and Y.~Bengio.
\newblock Fitnets: hints for thin deep nets.
\newblock In {\em Proc. ICLR}, 2015.

\bibitem{at:scheme}
S.~Zagoruyko and N.~Komodakis.
\newblock Paying more attention to attention: Improving the performance of
  convolutional neural networks via attention transfer.
\newblock In {\em Proc. ICLR}, 2017.

\bibitem{fsp:scheme}
J.~Yim, D.~Joo, J.~Bae, and J.~Kim.
\newblock A gift from knowledge distillation: Fast optimization, network
  minimization and transfer learning.
\newblock In {\em Proc. CVPR}, 2017.

\bibitem{jacobian:scheme}
S.~Srinivas and F.~Fleuret.
\newblock Knowledge transfer with jacobian matching.
\newblock In {\em Proc. ICML}, 2018.

\bibitem{firstasr:scheme}
J.~Li, R.~Zhao, T.~J. Huang, and Y.~Gong.
\newblock Learning small-size dnn with output-distribution-based criteria.
\newblock In {\em Proc. INTERSPEECH}, 2014.

\bibitem{chebotar-et-al:scheme}
Y.~Chebotar and A.~Waters.
\newblock Distilling knowledge from ensembles of neural networks for speech
  recognition.
\newblock In {\em Proc. INTERSPEECH}, pages 3439--3443, 2016.

\bibitem{watanabe-et-al:scheme}
S.~Watanabe, T.~Hori, J.~L.~Roux, and J.~R. Hershey.
\newblock Student-teacher network learning with enhanced features.
\newblock In {\em Proc. ICASSP}, pages 5275--5279, 2017.

\bibitem{lu-et-al:scheme}
L.~Lu, M.~Guo, and S.~Renals.
\newblock Knowledge distillation for small-footprint highway networks.
\newblock In {\em Proc. ICASSP}, pages 4820--4824, 2017.

\bibitem{fukuda-et-al:scheme}
T.~Fukuda, M.~Suzuki, G.~Kurata, S.~Thomas, J.~Cui, and B.~Ramabhadran.
\newblock Efficient knowledge distillation from an ensemble of teachers.
\newblock In {\em Proc. INTERSPEECH}, pages 3697--3701, 2017.

\bibitem{blending:scheme}
K.~J. Geras et~al.
\newblock Blending lstms into cnns.
\newblock In {\em Proc. ICLR Workshop}, 2016.

\bibitem{seq:scheme}
J.~H.~M. Wong and M.~J.~F. Gales.
\newblock Sequence student-teacher training of deep neural networks.
\newblock In {\em Proc. INTERSPEECH}, pages 2761–--2765, 2016.

\bibitem{seq2:scheme}
J.~H.~M. Wong, M.~J.~F. Gales, and Y.~Wang.
\newblock General sequence teacher–student learning.
\newblock {\em IEEE/ACM Transactions on Audio, Speech, and Language
  Processing}, 27(11):1725–--1736, 2019.

\bibitem{senior-et-al:scheme}
A.~Senior, H.~Sak, F.~C.~Quitry, T.~Sainath, K.~Rao, et~al.
\newblock Acoustic modelling with cd-ctc-smbr lstm rnns.
\newblock In {\em Proc. ASRU}, pages 604--609, 2015.

\bibitem{takashima-et-al:scheme}
R.~Takashima, S.~Li, and H.~Kawai.
\newblock An investigation of a knowledge distillation method for ctc acoustic
  models.
\newblock In {\em Proc. ICASSP}, pages 5809--5813, 2018.

\bibitem{takashima-et-al2:scheme}
R.~Takashima, S.~Li, and H.~Kawai.
\newblock Investigation of sequence-level knowledge distillation methods for
  ctc acoustic models.
\newblock In {\em Proc. ICASSP}, pages 6156--6160, 2019.

\bibitem{kurata2-et-al:scheme}
G.~Kurata and K.~Audhkhasi.
\newblock Improved knowledge distillation from bi-directional to
  uni-directional lstm ctc for end-to-end speech recognition.
\newblock In {\em Proc. SLT}, pages 411--417, 2018.

\bibitem{kurata-et-al:scheme}
G.~Kurata and K.~Audhkhasi.
\newblock Guiding ctc posterior spike timings for improved posterior fusion and
  knowledge distillation.
\newblock In {\em Proc. INTERSPEECH}, pages 1616--1620, 2019.

\bibitem{sak2015learning}
H.~Sak, A.~Senior, K.~Rao, O.~Irsoy, A.~Graves, F.~Beaufays, and J.~Schalkwyk.
\newblock Learning acoustic frame labeling for speech recognition with
  recurrent neural networks.
\newblock In {\em Proc. ICASSP}, pages 4280--4284. IEEE, 2015.

\bibitem{panayotov-et-al:scheme}
V.~Panayotov, G.~Chen, D.~Povey, and S.~Khudanpur.
\newblock Librispeech: an asr corpus based on public domain audio books.
\newblock In {\em Proc. ICASSP}, pages 5206--5210, 2015.

\bibitem{aishell:scheme}
J.~Du, X.~Nai, X.~Liu, and H.~Bu.
\newblock Aishell-2: transforming mandarin asr research into industrial scale.
\newblock {\em arXiv preprint arXiv:1808.10583}, 2018.

\bibitem{hybrid_ctc_attention:scheme}
S.~Watanabe, T.~Hori, S.~Kim, J.~R. Hershey, and T.~Hayashi.
\newblock Hybrid ctc/attention architecture for end-to-end speech recognition.
\newblock {\em IEEE Journal of Selected Topics in Signal Processing},
  11(8):1240–--1253, 2017.

\bibitem{vgg-asr:scheme}
Y.~Zhang, W.~Chan, and N.~Jaitly.
\newblock Very deep convolutional networks for end-to-end speech recognition.
\newblock In {\em Proc. ICASSP}, pages 4845–--4849. IEEE, 2017.

\bibitem{hori-et-al:scheme}
T.~Hori, S.~Watanabe, Y.~Zhang, and W.~Chan.
\newblock Advances in joint ctc-attention based end-to-end speech recognition
  with a deep cnn encoder and rnn-lm.
\newblock In {\em Proc. INTERSPEECH}, pages 949–--953, 2017.

\bibitem{vgg:scheme}
K.~Simonyan and A.~Zisserman.
\newblock Very deep convolutional networks for large-scale image recognition.
\newblock {\em arXiv preprint arXiv:1409.1556}, 2014.

\bibitem{las:scheme}
W.~Chan, N.~Jaitly, Q.~V. Le, and O.~Vinyals.
\newblock Listen, attend and spell: a neural network for large vocabulary
  conversational speech recognition.
\newblock In {\em Proc. ICASSP}, pages 4960–--4964. IEEE, 2016.

\bibitem{openseq2seq}
O.~Kuchaiev, B.~Ginsburg, I.~Gitman, V.~Lavrukhin, J.~Li, H.~Nguyen, C.~Case,
  and P.~Micikevicius.
\newblock Mixed-precision training for nlp and speech recognition with
  openseq2seq.
\newblock {\em arXiv preprint arXiv:1805.10387}, 2018.

\bibitem{espnet:scheme}
S.~Watanabe et~al.
\newblock Espnet: end-to-end speech processing toolkit.
\newblock In {\em Proc. INTERSPEECH}, pages 2207–--2211, 2018.

\bibitem{nemo:scheme}
O.~Kuchaiev et~al.
\newblock Nemo: a toolkit for building ai applications using neural modules.
\newblock {\em arXiv preprint arXiv:1909.09577}, 2019.

\bibitem{kingma-et-al:scheme}
D.~P. Kingma and J.~Ba.
\newblock Adam: a method for stochastic optimization.
\newblock In {\em Proc. ICLR}, 2015.

\bibitem{ginsburg-et-al:scheme}
B.~Ginsburg, P.~Castonguay, O.~Hrinchuk, O.~Kuchaiev, V.~Lavrukhin, R.~Leary,
  J.~Li, H.~Nguyen, and J.~M. Cohen.
\newblock Stochastic gradient methods with layer-wise adaptive moments for
  training of deep networks.
\newblock {\em arXiv preprint arXiv:1905.11286}, 2019.

\bibitem{Heafield2011KenLMFA}
K.~Heafield.
\newblock Kenlm: faster and smaller language model queries.
\newblock In {\em Proc. EMNLP}, 2011.

\bibitem{bpe:scheme}
R.~Sennrich, B.~Haddow, and A.~Birch.
\newblock Neural machine translation of rare words with subword units.
\newblock In {\em Proc. ACL}, pages 1715–--1725, 2016.

\bibitem{adadelta:scheme}
M.~D. Zeiler.
\newblock Adadelta: an adaptive learning rate method.
\newblock {\em arXiv preprint arXiv:1212.5701v1}, 2012.

\bibitem{graves-jaitly:kl-one}
A.~Graves and N.~Jaitly.
\newblock Towards end-to-end speech recognition with recurrent neural networks.
\newblock In {\em Proc. ICML}, pages 1764--1772, 2014.

\bibitem{specaug:scheme}
D.~S. Park, W.~Chan, C.~Zhang, Y.~Chiu, B.~Zoph, E.~D. Cubuk, and Le.~Q. V.
\newblock Specaugment: a simple data augmentation method for automatic speech
  recognition.
\newblock In {\em Proc. INTERSPEECH}, pages 2613--2617, 2019.

\bibitem{garg-et-al:scheme}
A.~Garg, O.~Gowda, A.~Kumar, K.~Kim, M.~Kumar, and Kim C.
\newblock Improved multi-stage training of online attention-based
  encoder-decoder models.
\newblock In {\em Proc. ASRU}, pages 70–--77, 2019.

\bibitem{audhkhasi-et-al:scheme}
K.~Audhkhasi, B.~Ramabhadran, G.~Saon, and D.~Picheny, M.~Nahamoo.
\newblock Direct acoustics-to-word models for english conversational speech
  recognition.
\newblock In {\em Proc. INTERSPEECH}, pages 959–--963, 2017.

\end{thebibliography}
\end{document}